\documentclass[12pt,AMA,STIX1COL]{WileyNJD-v2}

\articletype{RESEARCH ARTICLE}%
\usepackage{setspace}
\usepackage{amsmath, amssymb}
\usepackage{mathbbol}
\usepackage{bm} % Math packages
\def\*#1{\mathbf{#1}}
\def\^#1{\amsmathbb{#1}}
\def\##1{\mathbb{#1}}
\DeclareSymbolFontAlphabet{\amsmathbb}{AMSb}%

\received{26 April 2016}
\revised{6 June 2016}
\accepted{6 June 2016}

\begin{document}

\title{Scalar on time-by-distribution regression and its application for modelling  associations between daily-living physical activity and cognitive functions in Alzheimer’s Disease}

\author[1]{Rahul Ghosal*}

\author[2]{Vijay R. Varma}

\author[3]{Dmitri Volfson}
\author[4]{Jacek Urbanek}
\author[5,7,8]{Jeffrey M. Hausdorff}
\author[6]{Amber Watts}
\author[1]{Vadim Zipunnikov}

\authormark{Ghosal \textsc{et al}}

\address[1]{Department of Biostatistics, Johns Hopkins Bloomberg School of Public Health, Baltimore, Maryland USA }

\address[2]{National Institute on Aging (NIA), National Institutes of Health (NIH), Baltimore, Maryland, USA}

\address[3]{Neuroscience Analytics, Computational Biology, Takeda, Cambridge, MA, USA}

\address[4]{Department of Medicine, Johns Hopkins University School of Medicine, Baltimore Maryland, USA}

\address[5]{Center for the Study of Movement, Cognition and Mobility, Neurological Institute, Tel Aviv Sourasky Medical Center, Tel Aviv, Israel}

\address[6]{Department of Psychology, University of Kansas, Lawrence, KS, USA}

\address[7]{Department of Physical Therapy, Sackler Faculty of Medicine, and Sagol School of Neuroscience, Tel Aviv University, Tel Aviv, Israel}

\address[8]{Rush Alzheimer’s Disease Center and Department of Orthopedic Surgery, Rush University Medical Center, Chicago, USA}

\corres{*Rahul Ghosal \email{rghosal@ncsu.edu}}

%\presentaddress{This is sample for present address text this is %sample for present address text}

\abstract[Summary]{Wearable data is a rich source of information that can provide deeper understanding of links between human behaviours and human health. Existing modelling approaches use wearable data summarized at subject level via scalar summaries using regression techniques, temporal (time-of-day) curves using functional data analysis (FDA), and distributions using distributional data analysis (DDA). We propose to capture temporally local distributional information in wearable data using subject-specific time-by-distribution (TD) data objects. Specifically, we propose scalar on time-by-distribution regression (SOTDR) to model associations between scalar response of interest such as health outcomes or disease status and TD predictors. We show that TD data objects can be parsimoniously represented via a collection of time-varying L-moments that capture distributional changes over the time-of-day. The proposed method is applied to the accelerometry study of mild Alzheimer’s disease (AD). Mild AD is found to be significantly associated with reduced maximal level of physical activity, particularly during morning hours. It is also demonstrated that TD predictors attain much stronger associations with clinical cognitive scales of attention, verbal memory, and executive function when compared to predictors summarized via scalar total activity counts, temporal functional curves, and quantile functions. Taken together, the present results suggest that the SOTDR analysis provides novel insights into cognitive function and AD.}

\maketitle

%\footnotetext{\textbf{Abbreviations:} ANA, anti-nuclear antibodies; %APC, antigen-presenting cells; IRF, interferon regulatory factor}

\section{INTRODUCTION}\label{sec1}
Wearables are electronic sensors which can be worn as accessories and provide almost real-time continuous streams of user-specific physiological data such as minute-level step counts, heart rate (beats per minute and ECG), brainwave (EEG), and many others. This rich source of information can be analyzed for deeper understanding of human behaviours and their influence on human health and disease. For example, wearable physical activity (PA) monitors provide continuous and objective measurements of PA of individuals in their free-living environment \citep{karas2019accelerometry}.  The diverse applications of wearable data in biosciences include studies of aging \citep{varma2017re,schrack2014assessing}, circadian rhythms \citep{xiao2015quantifying}, estimation of gait parameters and their application in clinical trials \citep{urbanek2018validation,gait2020vr}, comparing patterns and intensity of physical activity between different clinical groups\citep{varma2017daily, watts2016intra} among many others. 

In many epidemiological and  clinical studies, wearable data is summarized via scalar summaries such as total log activity count (TLAC) \citep{varma2017re}, minutes of moderate-to-vigorous-intensity physical activity (MVPA) \citep{varma2017re,bakrania2017associations}, active-to-sedentary transition probability (ASTP) \citep{di2017patterns, schrack2019active} and others. Scalar summaries, although useful for a particular problem of interest, can often ignore temporal and/or distributional information in continuous streams of data. Temporal or time-of-day information in wearable data can be accounted for using functional data analysis (FDA) approaches that treat wearable data streams as functional observations recorded over 24 hours \citep{morris2006using, xiao2015quantifying, goldsmith2016new, cui2020additive}. 
Temporal effects of scalar predictors on physical activity can be captured via function-on-scalar regression and generalized multilevel function-on-scalar regression model \citep{goldsmith2015generalized}.
Scalar outcomes of interest, e.g., health or disease status can be modelled via scalar-on-function regression models \citep{reiss2017methods, leroux2019organizing} using diurnal physical activity curves as functional predictors typically averaged across the days of observation. 

Distributional information in wearable data can be accounted for using distributional data analysis (DDA). Distributions can be encoded via subject-specific histograms \citep{augustin2017modelling}, subject-specific quantile functions \cite{yang2020quantile, gait2020rv,matabuena2021distributional} or subject-specific densities \citep{kokoszka2019forecasting, tang2020differences, matabuena2020glucodensities}. The quantile-function based representation of information in wearable data allows us to model not just mean, but all other quantile-based distributional aspects of wearable data such as variability, skewness, and others. Ghosal et al. \citep{gait2020rv} developed a scalar-on-quantile function regression framework (SOQFR) for modelling scalar outcomes of interest based on subject specific quantile functions of wearable data. Matabuena and Petersen \citep{matabuena2021distributional} used quantile-function representation for NHANES (2003-2006) accelerometer data
to predict health outcomes using survey weighted nonparametric regression models. 
In this article, we propose to use time-by-distribution data objects that capture temporally local distributional information in the user-specific wearable data.  In previous work, Horvath et al \citep{horvath2020monitoring} proposed a statistical testing framework for detecting a change in a sequence of distributions, but the distributions were coming from the same unit (monthly financial returns of the same stock). Sharma et al \citep{sharma2020trajectories} considered distributions over space by time domain and modelled the change over time as linear with respect to the Wasserstein distance. Our approach is different in modelling subject-specific time-by-distribution objects that may have non-linear effects on the outcome. Note that
two different subjects could have markedly different diurnal patterns of activity but similar distributions, the proposed time-by-distribution PA metric, on the other hand, is more general and captures both these aspects jointly. The bivariate functional summaries of PA can be further used in penalized scalar-on-function regression (SOFR) \citep{marx2005multidimensional} for modelling scalar response of interest, such as cognitive outcomes in our motivating study. We use a penalized bivariate SOFR approach, which simultaneously identifies time of the day and quantiles of the subject-specific PA distribution, which are associated with cognitive status and cognitve functions. In addition, the bivariate time-by-distribution encoding of PA is shown to be equivalent to an alternative and parsimonious representation of wearable data in terms of diurnal time-varying L-moments \citep{hosking1990moments}, offering both temporal and distributional interpretation.

We are motivated by application of wearable data in the study of Alzheimer’s Disease (AD) and cognitive performance among older adults. AD is one of the most rapidly growing neurodegenerative diseases in the world. The high prevalence of AD and AD-related death in developed countries can be partially attributed to low levels of physical activity (PA) and sedentary lifestyles \citep{gronek2019physical}. In the absence of any currently existing cure for AD, there is growing interest in identifying cost effective biomarkers for early identification of risk for AD. Non-invasive, cost-efficient biomarkers are essential for improving early diagnosis of AD \citep{zvvevrova2018alzheimer}. “Digital” biomarkers from sensor and mobile/wearable devices \citep{kourtis2019digital} offers an alternative to existing fluid and imaging markers and there is a growing body of evidence which suggests PA changes might precede clinical manifestation of the disease itself. Physical activities, including activities of everyday living (ADLs), are dependent on mobility and cognitive functioning. Several prospective longitudinal studies have identified physical inactivity as a risk factor for dementia \citep{larson2006exercise,andel2008physical,geda2010physical,buchman2012total}. Older adults generally spend most of their waking time in sedentary activities \citep{harvey2013prevalence} and individuals with Alzheimer’s disease (AD) have been found to be even less active in previous studies \citep{watts2013metabolic}. 

%In a study by Varma and Watts\cite{varma2017daily}, mild AD was found to be associated with reduced moderate-intensity physical activity, reduced peak activity but not with increased sedentary activity or reduced low-intensity physical activity.

In our motivating study by Varma and Watts \citep{varma2017daily}, physical activity was monitored continuously for seven days using body-worn accelerometers in older adults with mild AD and cognitively normal controls (CNC). Mild AD was found to be associated with reduced moderate-intensity physical activity, reduced peak activity but not with increased sedentary activity or reduced low-intensity physical activity. Although prior research have focused on exploring effects of mild AD on diurnal patterns of PA \citep{varma2017daily} and on average or IIV (intra-individual variability) of PA across days \citep{watts2016intra}, we are interested in whether joint modelling of physical activity using temporally local distributional information can be used to differentiate between CNC and mild AD and participants and explain cognitive performance. 

The article is organized as follows. In Section \ref{sec2}, we present the background of our motivating study. In section \ref{mf}, we present our modelling framework and illustrate some existing approaches for modelling scalar response of interest e.g., cognitive outcomes based on scalar, temporal and distributional summary of wearable data. In Section \ref{method}, we introduce the time-by-distribution PA metric and illustrate an estimation approach using penalized bivariate scalar-on-function regression. In addition, an alternative formulation of the problem in terms of subject-specific and diurnal time-varying  L-moments of physical activity data is also provided. In Section \ref{appsec}, we demonstrate applications of the proposed method in the Alzheimer's disease (AD) study and provide comparison with existing approaches. Section \ref{sec6} concludes with a discussion of the findings, limitations and some possible extensions of this work.

\section{Motivating Study}\label{sec2}
\subsection{Study Participants}
Mild AD and cognitively normal control (CNC) participants were recruited by the University of Kansas Alzheimer’s Disease Center Registry (KU-ADC). The study protocol was approved by the KU Medical Center Institutional Review Board.  Detailed description of recruitment and evaluation of participants in the KU-ADC have been previously reported in Graves at al. \cite{graves2015open}. All participants received annual cognitive and clinical examinations, and experienced clinicians trained in dementia assessment provided consensus diagnoses (see a subsection on cognitive status below for more details). The study sample consisted of individuals with mild AD, defined as a clinical dementia rating (CDR; \citep{morris1991clinical}) scale scores of 0.5 (very mild) or 1 (mild), and control participants, defined as a CDR score of 0. A total of 100 community dwelling older adults with and without mild AD were recruited. Out of them, N=92 had valid actigraphy data (n = 39 mild AD; n = 53 controls). Descriptive summaries of participant demographics are displayed in Table \ref{tab:my-table1}. Age, sex, and years of formal education were reported by either the participant or study partner. The details about other measures are provided in Graves et al \cite{graves2015open}.

\begin{table}[ht]
\centering
\caption{Summary statistics for the complete, AD and CNC samples. No statistical difference between the AD and CNC groups are observed across age, BMI, or V$0_2$ max. However, AD group had a smaller percentage of females ($28.2$ vs $69.8$ for CNC) and lower education ($15.5$ years vs $17.3$ years for CNC).}
\label{tab:my-table1}
\begin{tabular}{cccccccc}
\hline
Characteristic     & \multicolumn{2}{c}{Complete sample} & \multicolumn{2}{c}{AD} & \multicolumn{2}{c}{CNC} & P value         \\ \hline
                   & Mean/Freq           & SD             & Mean/Freq     & SD      & Mean/Freq       & SD        &                 \\ \hline
Age                & 73.36               & 7.11           & 73.59         & 7.92    & 73.19           & 6.53      & 0.797           \\ \hline
\% Female          & 52.17                  & N/A            & 28.20         & N/A     & 69.81           & N/A       & \textless 0.001 \\ \hline
Years of edu & 16.56               & 3.24           & 15.53         & 2.77    & 17.32           & 3.38      & 0.0064          \\ \hline
BMI                & 26.78               & 4.52           & 27.28         & 5.04    & 26.42            & 4.11      & 0.3892          \\ \hline
VO2 max            & 21.99               & 5.34          & 21.61         & 5.24    & 22.24           & 5.43      & 0.592          \\ \hline
\end{tabular}
\end{table}

\subsection{Physical activity}
Activity counts were produced by a GT3x+ tri-axial accelerometer. A detailed description of accelerometry measurement can be found in \cite {varma2017daily}.  Briefly, the GT3x+ (Pensacola FL; Actigraph, 2012; 30 Hz sampling rate) is a triaxial accelerometer validated across a range of community dwelling older adults. The accelerometer was placed on the dominant hip of the participants via elastic belt and the participants were instructed to wear the device for 24 hours a day for seven days. Activity counts, collected every second from medio-lateral (ML; front-to-back), antero-posterior (AP; side-to-side), and vertical (VT; rotational) axes were quantified into a single tri-axial composite metric known as vector magnitude \citep{actigraph2012actilife}, calculated as $VM=\sqrt{ML^2+AP^2+VT^2}$. Average vector magnitude was then computed by aggregating VM (averaging) for each second into minute level activity.

\subsection{Cognitive status and psychometric test battery}
 Cognitive status of the participants were determined through consensus diagnosis by trained clinicians using comprehensive clinical research evaluations and a review of medical records following NINCS-ADRDA criteria \citep{mckhann1984clinical}.  Cognitive tests were administered by a trained psychometrician. The cognitive test battery included tests of verbal memory (Wechsler Memory Scale (WMS)–Revised Logical Memory I and II, Free and Cued Selective Reminding Task), attention (Digits Forward and Backward, Wechsler Adult Intelligence Scale (WAIS) subscale Letter– Number Sequencing) and executive function (Digit Symbol Substitution Test, and Stroop Color–Word Test (interference score), Trail Making Test Part B, and Category Fluency). Composite scores for each domain (verbal memory (VM), attention (ATTN), and executive function (EF)) were derived using confirmatory factor analysis (CFA), a flexible approach of summarizing multiple cognitive scores into empirically and theoretically justified components. Scores were standardized to the mean performance of CNC participants. Additional information on the CFA derived factor scores can be found in Varma et al. \citep{gait2020vr}.
 
 \section{Modelling Frameworks}
 \label{mf}
 Suppose, we have minute-level wearable observations such as activity counts or the number of steps per minute denoted by $X_{ij}(t) \hspace{1 mm}$ for subject $i=1,\ldots,n$, on $j$-th day,  $j=1,\ldots,n_i$, at time $t$, $t=1,2,\ldots,1440$. We denote by $Y_i$ a scalar outcome of interest such as a cognitive status or a score on psychometric test that can be continuous or discrete and we assume it comes from an exponential family. We also denote by $\*Z_i$ a vector of covariates. In this section, we review three existing modelling approaches that relates $Y_i$ and $X_{ij}(t)$ including simple regression using scalar summaries of wearable observations,  functional data analysis of temporal (time-of-day) curves, and distributional data analysis using subject-specific quantile functions.

 \subsection{Regression using subject-specific scalar summaries}
 
\begin{figure}[ht]
\begin{center}
\begin{tabular}{ll}
\includegraphics[width=.5\linewidth , height=.48\linewidth]{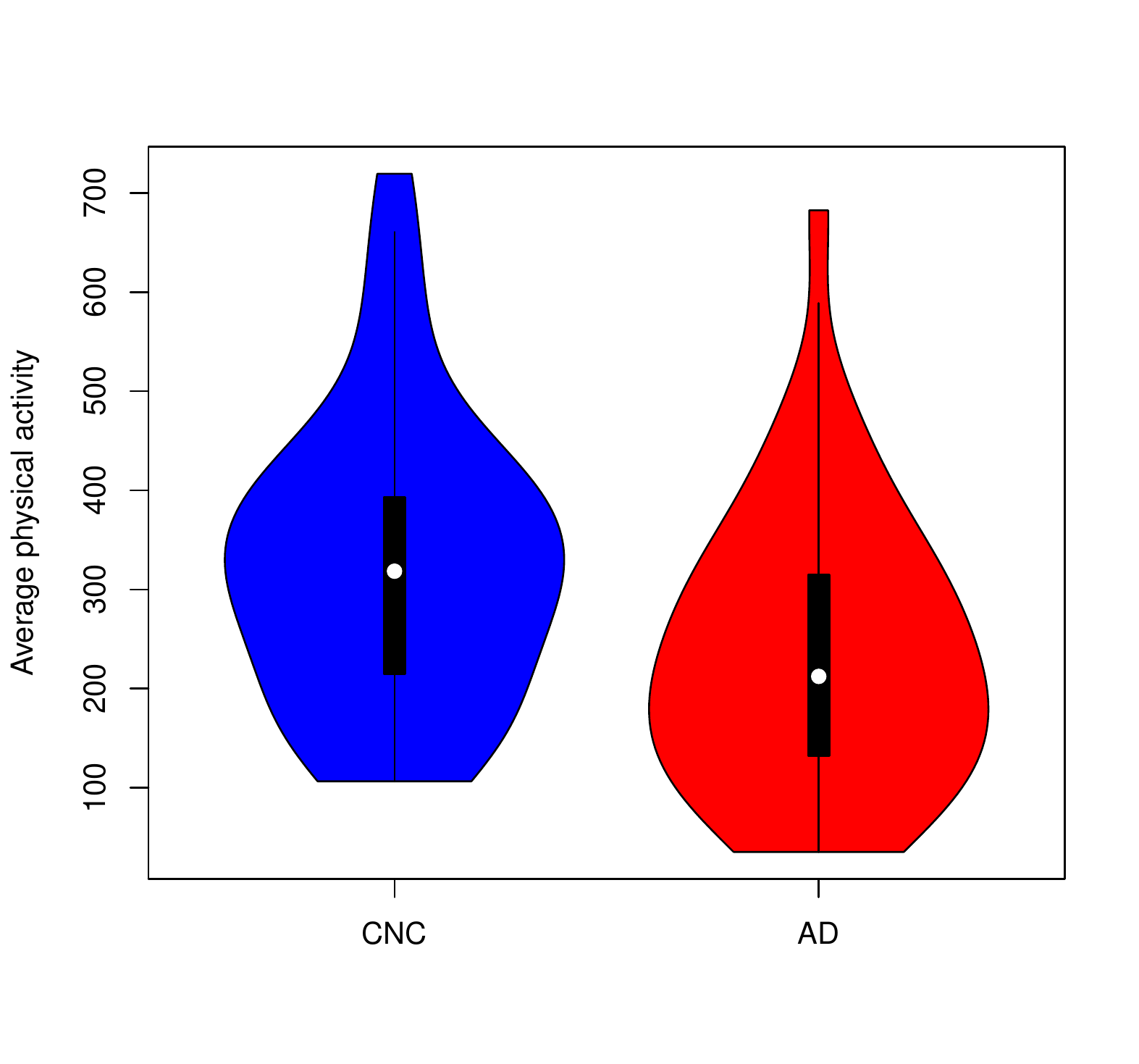} &
\includegraphics[width=.5\linewidth , height=.48\linewidth]{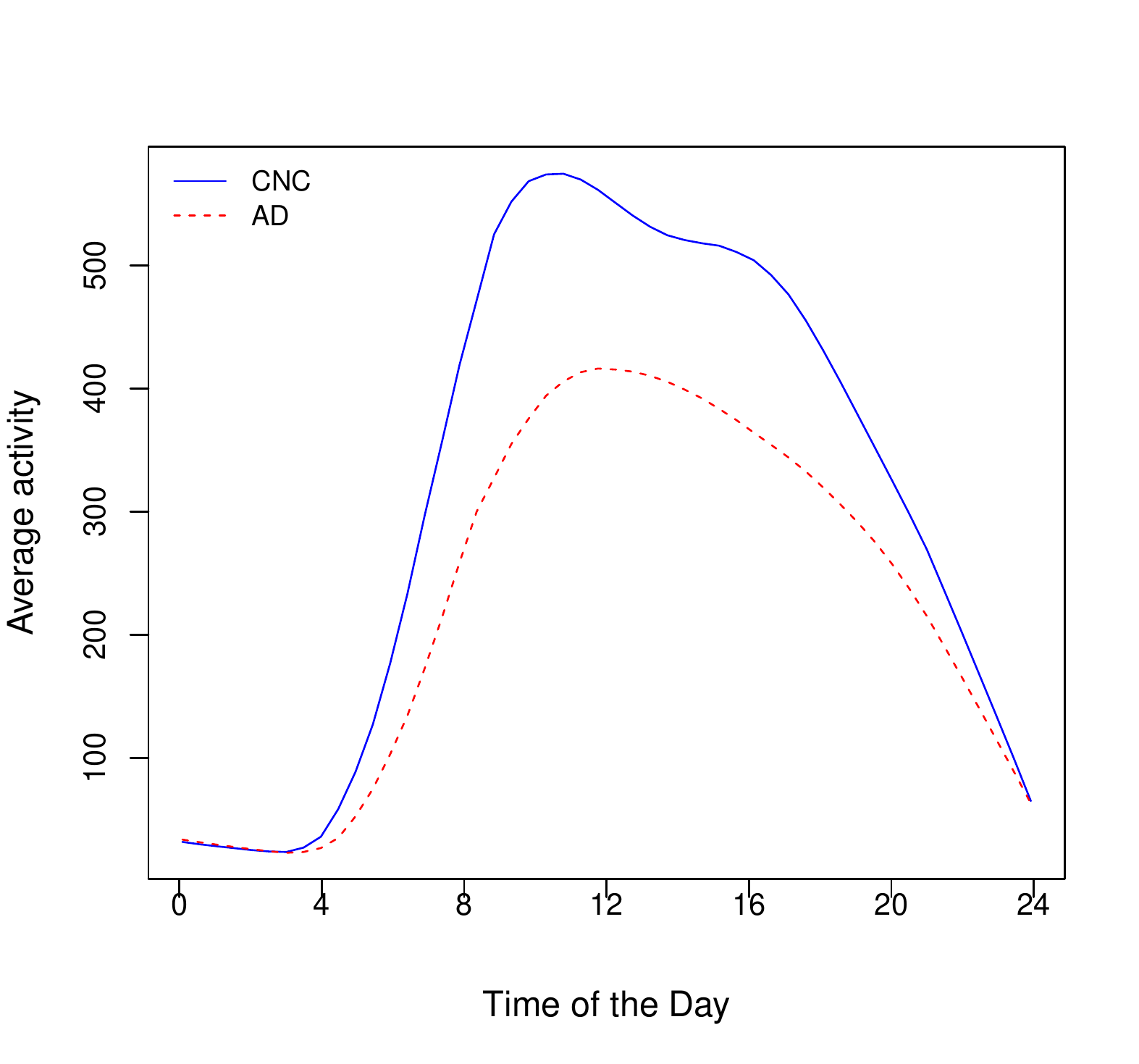}\\
\includegraphics[width=.5\linewidth , height=.48\linewidth]{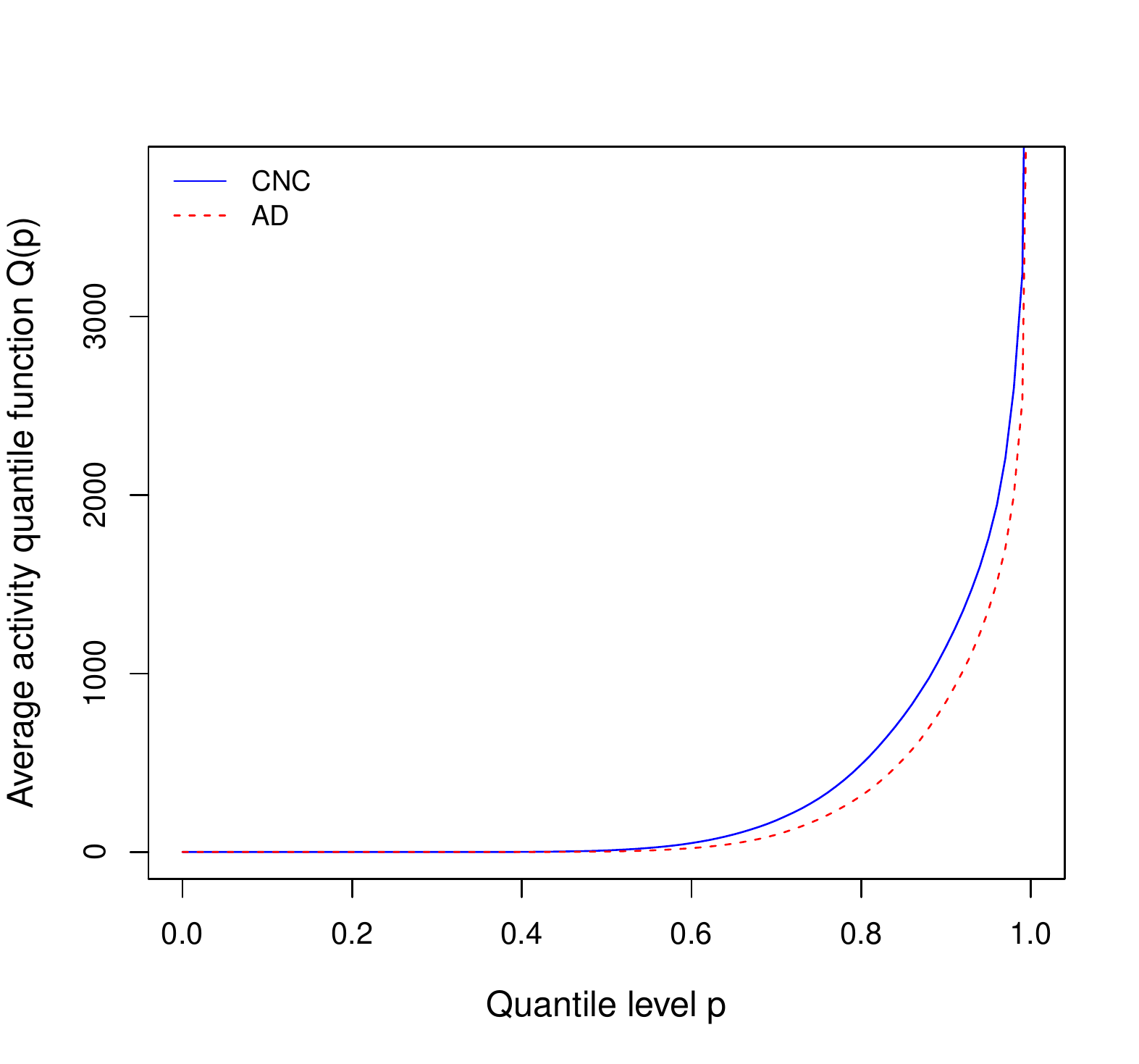}& \\
\end{tabular}
\end{center}
\caption{top left: violin plot of subject-specific averages for CNC and AD participants. top right: smoothed diurnal activity profiles averaged across CNC (blue) and AD (red) participants. Bottomleft: average quantile functions of physical activity for AD and CNC participants.}
\label{fig:fig1c}
\end{figure}

In this approach, the scalar response variable $Y_i$ is modelled via the subject- specific scalar summary of wearable observations aggregated across all times and days. Examples include a  mean as a measure of tendency, a standard deviation as a measure of variability, minutes spent in activities of certain intensity such as light or moderate-to-vigorous, and others. For example, subject-specific average activity count $\bar{X}_i=\frac{1}{1440n_i} \sum_{j=1}^{n_i}\sum_{t=1}^{1440}X_{ij}(t)$. The top left panel of Figure \ref{fig:fig1c} displays the distribution of subject-specific averages for CNC (blue) and AD (red) groups in our study.
We observe that participants with AD on average, have a lower mean physical activity level compared to CNC. There is also a significant overlap between the two distributions and they are not clearly separable using this PA metric. To formally model this, a generalized linear model (GLM) can be used 
\begin{eqnarray}
E(Y_i|\bar{X}_i)=\mu_i, \hspace{2 mm}
g(\mu_i)=\alpha+\*Z_i^T\bm\gamma + \bar{X}_i\beta\label{sofr1},
\end{eqnarray}
where a scalar regression coefficient $\beta$ represents the effect of average PA on the mean of the response of interest $Y_i$ adjusted for covariates $\*Z_i$ and $g(\cdot)$ is a known link function (e.g., $logit$ or identity).

\subsection{Functional data analysis of subject-specific temporal curves}

Functional data analysis (FDA) allows us to model temporal aspects in wearable observations $X_{ij}(t)$. To derive subject-specific diurnal minute-level curves, one may average wearable observations across all days at each time-point $t=1,2,\ldots,1440$ as $X_i(t)=\frac{1}{n_i} \sum_{j=1}^{n_i}X_{ij}(t)$. The top right panel of Figure \ref{fig:fig1c}  displays average smoothed diurnal activity profiles for CNC (blue) and AD (red) groups. It can be noticed that the curve for mild-AD group have a unimodal diurnal shape, compared to a bimodal shape for CNC, and the largest difference between the two groups appears to be in the morning and in the afternoon (during the second peak for CNC). Similar observations were also made by Varma and Watts \cite{varma2017daily} during their analysis of this data. To formally model the association with functional predictors, scalar-on-function regression (SOFR)\citep{reiss2017methods} can be used as follows
\begin{eqnarray}
E(Y_i|X_i(t))=\mu_i, \hspace{2 mm}
g(\mu_i)=\alpha+\*Z_i^T\bm\gamma + \int_{T}X_i(t)\beta(t)dt\label{sofr2}, 
\end{eqnarray}
where the functional regression coefficient $\beta(t)$ captures the time-varying effect of the diurnal curve $X_i(t)$ on the response $Y_i$ and $T=(0,24)$ is the daily 24 hour window.  Note that, the average subject-specific PA can be estimated back from the diurnal profile $X_i(t)$ as $\bar{X}_i=\int_T X_i(t)dt$, therefore for a constant functional regression coefficient $\beta(t)=\beta$, one gets back the generalized linear model (\ref{sofr1}) for scalar predictors from model (\ref{sofr2}).

\subsection{Distributional data analysis using subject-specific quantile functions}
Distributional data analysis can capture and model distributional aspect of wearable observations via subject-specific pdfs, CDFs, or quantile functions \citep{gait2020rv}. If we ignore the temporal information by suppressing the time index $t$, we can denote by $X_{ik}$, $k=1,\ldots,m_i$, all wearable observations for subject $i$. We assume $X_{ik}$ follow the same  subject-specific distribution defined by subject-specific cumulative distribution function $F_i(x)$, where $F_i(x)=P(X_{ik}\leq x)$. Then, we can define the subject-specific quantile function $Q_i(p)= \inf\{x: F_i(x)\geq p\}$. The subject-specific quantile function characterizes the distribution of wearable observations for a specific subject. The subject-specific cdf can be estimated via its empirical counterpart $\hat{F}_i(x)=\frac{1}{m_i}\sum_{k=1}^{m_i}I(X_{ik}\leq x)$ and subject-specific quantile function can be  estimated as $\hat{Q}_i(p)=\hat{F}_i^{-1}(p)$. 
In this paper, we use the following estimator of quantile functions via a linear interpolation of the order statistics \citep{parzen2004quantile}: 
$$\hat{Q}(p)=(1-w)X_{([(n+1)p])}+wX_{([(n+1)p]+1)},$$
where $X_{(1)}\leq X_{(2)}\leq \ldots,X_{(n)}$ are the corresponding order statistics from a sample $(X_1,X_2,\ldots, X_n)$ and $w$ is a weight satisfying $(n+1)p=[(n+1)p]+w$. Note that the subject-specific average of wearable observations $X_{ij}(t)$ can be also estimated from the subject-specific quantile function as $\bar{X}_i=\int_{0}^{1} {Q}_i(p)dp$. 

The bottom left panel of Figure \ref{fig:fig1c} displays the average quantile functions of physical activity for the CNC and AD groups. A reduced capacity of physical activity can be observed for the AD samples compared to CNC across upper quantile levels such as $p > 0.75$. 
Following the approach of Ghosal et al. \cite{gait2020rv}, the subject-specific quantile functions of PA can be used for modelling $Y_i$ using scalar-on-function regression (SOFR) (\ref{sofr3})  adjusted for $\*Z_i$. SOFR model is as follows
\begin{eqnarray}
E(Y_i|Q_i(p))=\mu_i, \hspace{2 mm}
g(\mu_i)=\alpha+\*Z_i^T\bm\gamma + \int_{0}^{1}Q_i(p)\beta(p)dp\label{sofr3},
\end{eqnarray}
where the functional regression coefficient $\beta(p)$ captures the distributional effect of the PA quantile function $Q_i(p)$ on the response of interest $Y_i$. In the case $\beta(p)=\beta$, a constant, one again get back the generalized linear model (\ref{sofr1}) from model (\ref{sofr3}), since $\bar{X}_i=\int_{0}^{1} Q_i(p)dp$.

Ghosal et al. \citep{gait2020rv} re-represented SOFR model for quantile function predictors via L-moments  \citep{hosking1990moments}. L-moments are defined as the expectation of a linear combination of order statistics. In particular, the $r$-th order L-moment of a random variable $X$ is defined as 
\begin{equation*}
    L_r=r^{-1}\sum_{k=0}^{r-1}(-1)^k {r-1 \choose k} E(X_{r-k:r})\hspace{3mm} r=1,2,\ldots,
\end{equation*}
where $X_{1:n}\leq X_{2:n}\leq \ldots \leq X_{n:n}$ denote the order statistics of a random sample of size $n$ drawn from the distribution of $X$. 
The first order L-moment, $L_1$, equals the traditional mean. The second order L-moment, $L_2 = 1/2E(X_{2:2}-X_{1:2})$, represents a robust measure of scale, and equals exactly a half of Gini-coefficient or mean absolute difference. The third and fourth order L-moments, $L_3 = 1/3E(X_{3:3}-2X_{2:3}+X_{1:3})$ and $L_4 = 1/4E(X_{4:4}-3X_{3:4}+3X_{2:4}-X_{1:4})$, capture higher-order distributional properties and normalized by $L_2$ can be interpreted similarly to traditional higher-order moments such as skewness and kurtosis. The main advantages of L-moments is the existence of all moments, if first moment exist, their uniqueness and robustness.  For SOFR Ghosal et al. \citep{gait2020rv} adapted an alternative representation of L-moments as projections of quantile functions on Legendre polynomial basis, given by
\begin{equation*}
    L_r = \int_0^1 Q(p)P_{r-1}(p)dp. 
\end{equation*}
Here $P_r(p)$ is the shifted Legendre polynomial (LP) of degree $r$ defined as
\begin{equation*}
    P_r(p) = \sum_{k=0}^r s_{r,k}p^r, \quad s_{r,k} = (-1)^{r-k}{r \choose k}{r+k \choose k} = \frac{(-1)^{r-k}(r+k)!}{(k!)^2(r-k)!}.
\end{equation*}  
The shifted Legendre polynomials form an orthogonal basis of $L_2[0,1]$. Using the LP decomposition for subject-specific quantile functions $Q_i(p) \approx \sum_{k=1}^K(2k-1)L_{ik}P_{k-1}(p)$ and $\beta(p)=\sum_{k=1}^K\beta_kP_{k-1}(p)$, SOFR model can be reduced to a GLM as $g(\mu_i)=\alpha+\*Z_i^T\bm\gamma + \int_{0}^{1}Q_i(p)\beta(p)dp = \alpha+\*Z_i^T\bm\gamma + \sum_{k=1}^K \beta_kL_{ik}$. This representation of SOFR via L-moments provides both the functional  interpretation of significance of $Q_i(p)$ via $\beta(p)$ and the distributional interpretation in terms of the significance of specific L-moments via $\beta_k$.

\section{Scalar on time-by-distribution regression}
\label{method}
In this section, we propose to capture temporally local distributional information in wearable observations
using subject-specific time-by-distribution data objects. We develop scalar on time-by-distribution regression (SOTDR) and show how two-way TD data objects can be parsimoniously represented via a collection of time-varying L-moments that capture distributional changes over the time-of-day.

\subsection{SOTDR via time-by-distribution data objects}
\label{method1}
We propose quantile-based time-by-distribution data objects that  capture the temporally local distributional aspects of wearable observations. The quantile-based time-by-distribution data object is then defined as 
\begin{equation*}
    Q_i(t,p)= \textit{p-th quantile of \{$X_{ij}(s)\}_{j=1}^{n_i}$, $s \in (t-h,t+h)$}.
\end{equation*}
Here $2h$ is the window length around time $t$. Note that $Q_i(t,p)$ is a bivariate functional summary of subject-specific observational data. For each fixed $t$ (time of the day), it provides distributional encoding as a function of quantile-level $p$, e.g., $Q_i(t,\cdot)$ is a quantile function for each $t$. For each fixed $p$, $Q_i(\cdot,p)$ captures the diurnal pattern of the $p$-th quantile level of wearable observations as a function of time $t$. Note that the subject-specific average PA can be again be estimated back aggregating the bivariate time-by-distribution data objects as $\bar{X}_i=\int_{T}\int_{0}^{1} {Q}_i(t,p)dpdt$. For the analysis presented in this paper, we fix total window length $2h=10$ minutes (i.e., $h=5$), but any other window lengths can be used as well. Since the sample considered in this study is highly sedentary \citep{watts2016intra}, a window length of 10 minutes still retains the diurnal patterns of PA without any significant loss of information.

\begin{figure}[ht]
\begin{center}
\begin{tabular}{ll}
\includegraphics[width=.5\linewidth , height=.45\linewidth]{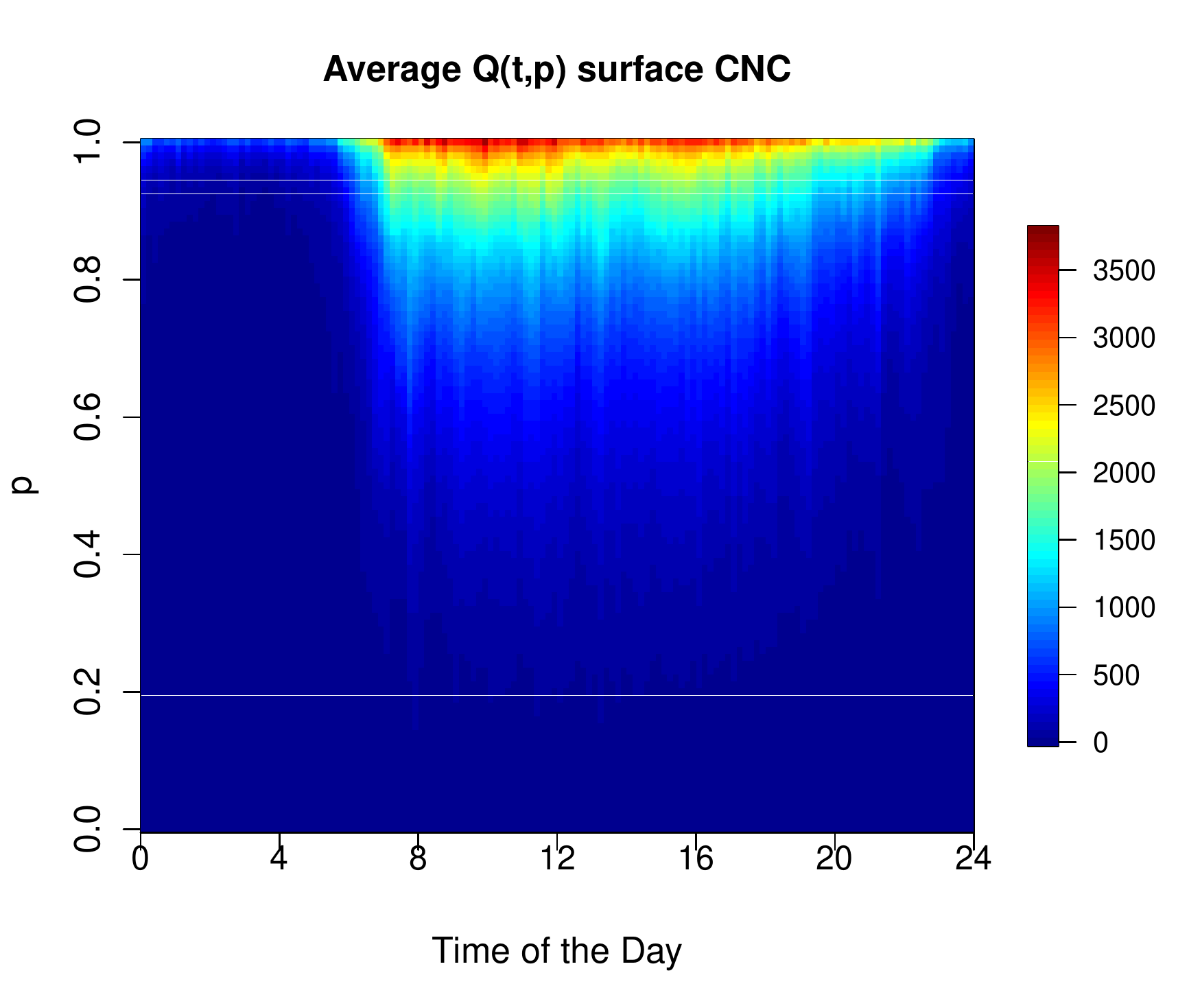} &
\includegraphics[width=.5\linewidth , height=.45\linewidth]{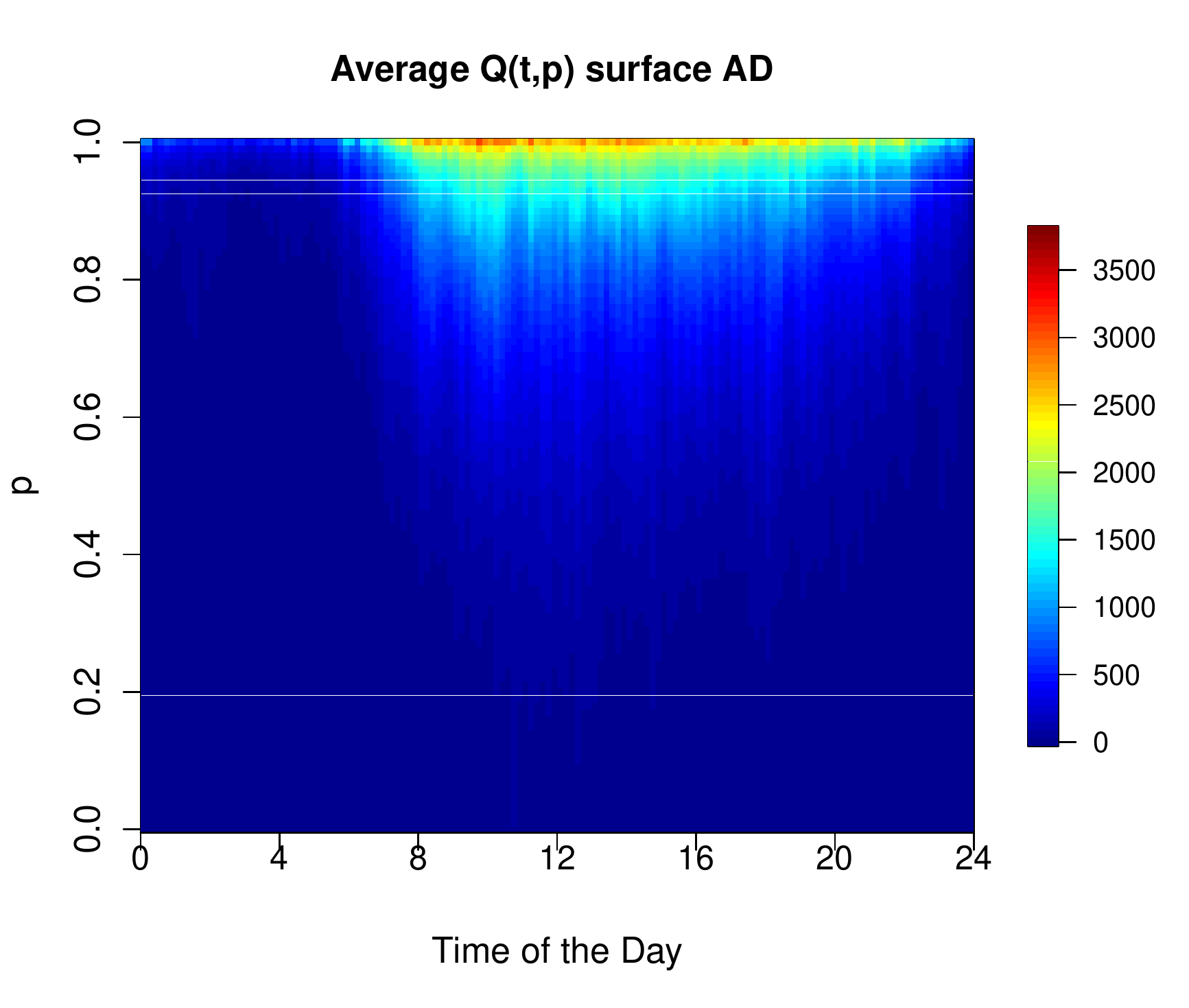}\\
\includegraphics[width=.5\linewidth , height=.45\linewidth]{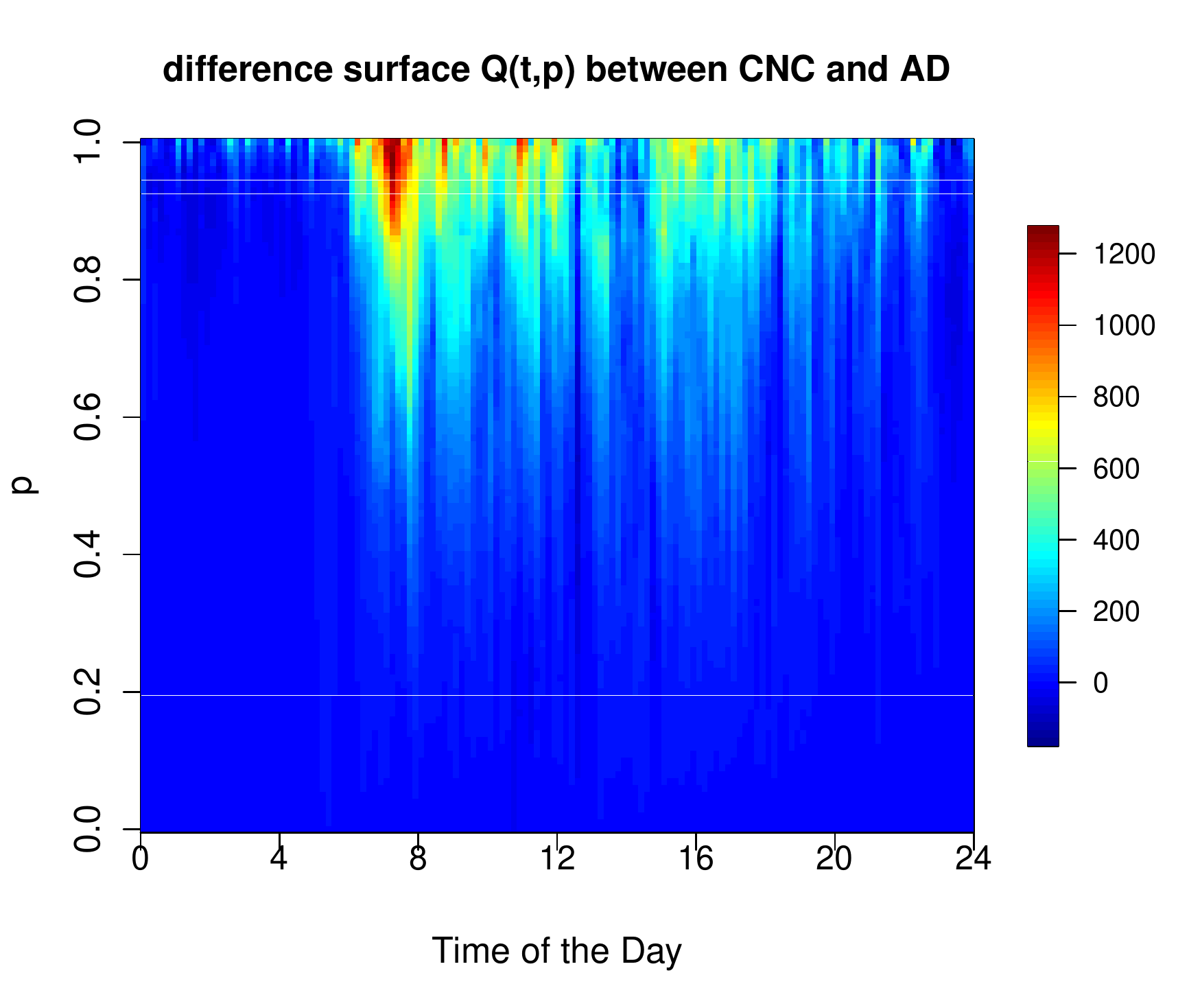} &
\includegraphics[width=.5\linewidth , height=.45\linewidth]{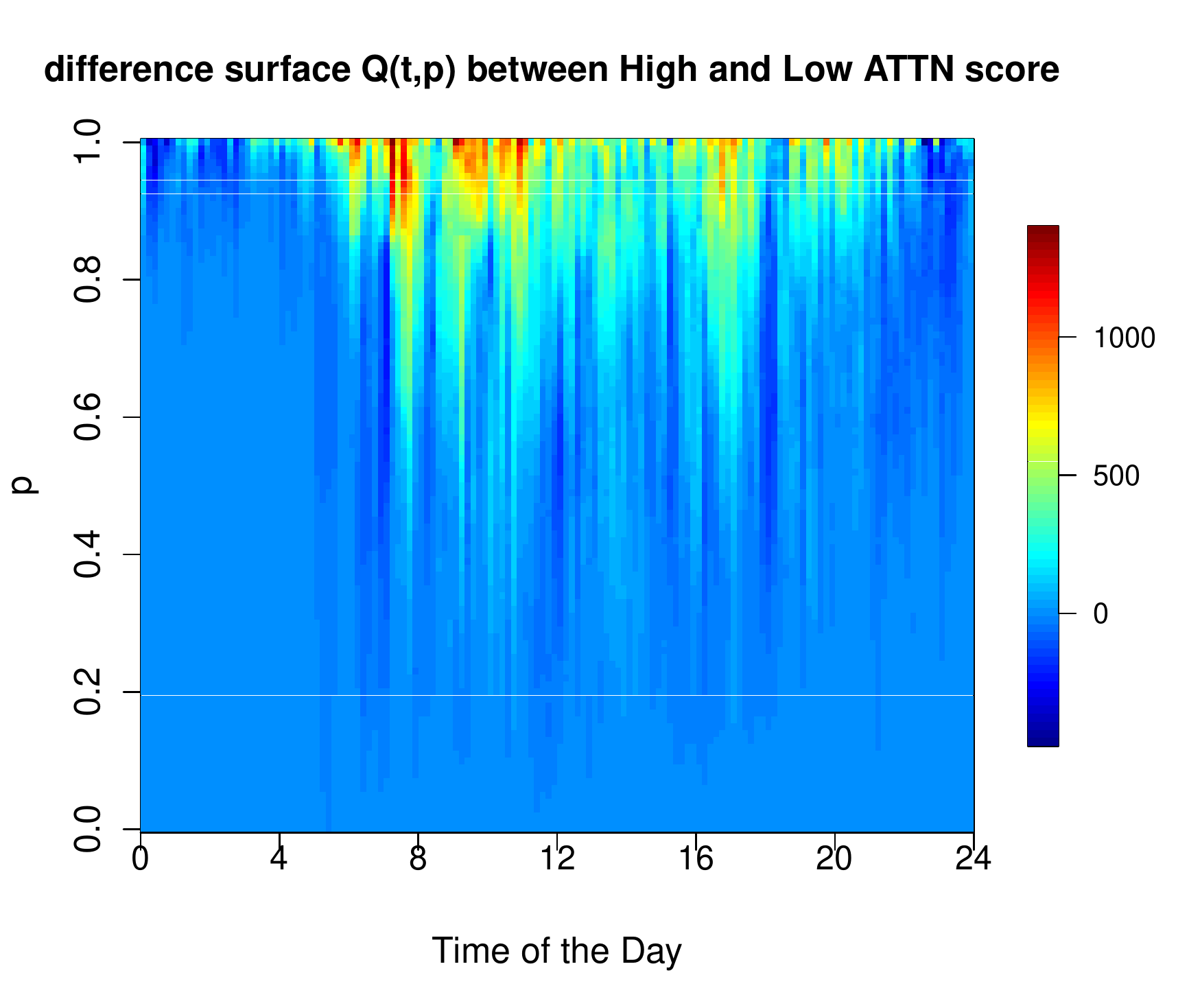}\\
\end{tabular}
\end{center}
\caption{The average bivariate time-by-distribution PA surface $Q_i(t,p)$ for CNC (top left) and AD (top right) groups. The difference between CNC and AD (bottomleft) and the difference between subjects with high (above $75\%$ percentile) and low (below $25\%$-percentile) of cognitive attention (ATTN) scores (bottomright). }
\label{fig:fig2}
\end{figure}

Figure \ref{fig:fig2} displays the heatmaps of average  time-by-distribution surfaces $Q_i(t,p)$ for CNC (top left) and AD (top right), the difference between them (bottomleft). One can see that the largest differences between the two groups exist during the morning (8 a.m-11 a.m) and in afternoon (3 p.m-5 p.m) across the upper quantile levels ($p>0.6$). At the bottomright panel of Figure \ref{fig:fig2} we plot the heatmap of difference in time-by-distribution surfaces $Q_i(t,p)$ between the participants with high (above $75\%$-percentile) and low (below $25\%$-percentile) cognitive score of attention (ATTN). Overall, TD encoding of physical activity is clearly more informative than just temporal or just distributional information from Figure \ref{fig:fig1c}.

To formally model the association of TD data objects with scalar response, we propose to use them as predictors in two-way scalar-on-function regression (SOFR) as follows: 
\begin{eqnarray}
E(Y_i|Q_i(t,p))=\mu_i, \hspace{2 mm}
g(\mu_i)=\alpha+\*Z_i^T\bm\gamma + \int_{0}^{1} \int_{T}Q_i(t,p)\beta(t,p)dtdp.\label{sofr}
\end{eqnarray}
Here $\beta(t,p)$ represents the bivariate functional regression coefficient that captures both the temporal and distributional effect of $Q_i(t,p)$ on the response of interest $Y_i$. As before, with the constant regression $\beta(t,p)=\beta$, the bivariate SOFR model (\ref{sofr}) reduces to the generalized linear model (\ref{sofr1}) for scalar predictors. The estimation approach of this model is discussed below.

\subsection*{Estimation of the time-by-distribution regression coefficient $\beta(t,p)$}
We follow a two-step estimation approach for the bivariate SOFR model (\ref{sofr}) in the paper. In step 1, we model the bivariate regression functional coefficient $\beta(t,p)$ using a tensor product of univariate cubic B-spline basis functions of both temporal and quantile level arguments, $t$ and $p$. Suppose, $\{B_{T,k}(t)\}^{K_0}_{k=1}$ and $\{B_{P,\ell}(p)\}^{L_0}_{\ell=1}$ are the set of known basis functions over $t$ and $p$, respectively.
Then, $\beta(t,p)$ is modelled as $\beta(t,p)=\sum_{k=1}^{K_0}\sum_{\ell=1}^{L_0}\theta_{k,\ell}B_{T,k}(t)B_{P,\ell}(p)$. Using this expansion model (\ref{sofr}) is reformulated as
\begin{eqnarray}
g(\mu_i)&=&\alpha+\*Z_i^T\bm\gamma + \sum_{k=1}^{K_0}\sum_{\ell=1}^{L_0}\theta_{k,\ell}\int_{T}\int_{0}^{1}Q_i(t,p)B_{T,k}(t)B_{P,\ell}(p)dtdp\notag\\
&=& \alpha+\*Z_i^T\bm\gamma +\*W_{i}^T\bm\theta, \label{FGAM2}
\end{eqnarray}
where we denote by $\*W_i$  the $K_0L_0$-dimensional stacked vectors of $\{\int_{0}^{1}Q_i(t,p)B_{T,k}(t)B_{P,\ell}(p)dtdp\}_{k=1,\ell=1}^{K_0,L_0}$  and $\bm\theta$ is the corresponding $K_0L_0$-dimensional vector of unknown basis coefficients $\theta_{k,\ell}$'s. Thus, the model (\ref{FGAM2}) can be seen as a GLM with subject specific predictors $\*W_{i}^{k,\ell}=\int_{0}^{1}Q_i(t,p)B_{T,k}(t)B_{P,\ell}(p)dtdp$. We use a penalized negative log-likelihood criterion with LASSO \citep{tibshirani1996regression} penalty on the coefficients, which simultaneously identifies the most informative time of the day and the quantile levels. This effectively helps to reduce the number of parameters (variables) in the model and allows a sparse representation of the functional regression coefficient $\beta(t,p)$, which is, especially, important when dealing with a relatively small sample size such as in our motivating application with $n  = 92$. The penalized negative log likelihood criterion is given by
\begin{equation}
S(\psi)=R(\alpha,\bm\gamma,\bm\theta)=  -2log L(\alpha,\bm\gamma,\bm\theta;Y_i,\*Z_i,\*W_i) + ||\bm\theta||_{1}.
\label{SOFR:cri}
\end{equation}
In step 2, the selected predictors $\*W_{i}^{k,\ell}$ (with non-zero coefficients) are used in the GLM (\ref{FGAM2}) without any penalization (this also overcomes penalization bias of LASSO) for inference.
The estimated regression coefficient function is then given  by $\hat{\beta}(t,p)=\sum_{k=1}^{K_0}\sum_{\ell=1}^{L_0}\hat{\theta}_{k,\ell}B_{T,k}(t)B_{P,\ell}(p)$ (note that $\hat{\theta}_{k,\ell}=0$ if $\*W_{i}^{k,\ell}$ is not selected in the first step).

\subsection{SOTDR-L: SOTDR via time-varying L-moments}
\label{methodLmom}
Following Ghosal et al. \citep{gait2020rv} who adapted L-moments to SOFR with quantile function predictors, we adapt L-moments to SOTDR by introducing subject-specific time-varying L-moments $L_{ir}(t)$ that depend on the time of the day $t$.  
Specifically, we define the diurnal time-varying $r$-th order L-moment for subject $i$ as   
\begin{equation*}
    L_{ir}(t)= \textit{r-th L-moment of \{$X_{ij}(s)\}_{j=1}^{n_i}$, $s \in (t-h,t+h)$}.
\end{equation*}
Here we again consider a window of total length $2h$ centered at time $t$. The diurnal time-varying $L_{ir}(t)$ curves capture the temporal change of the subject-specific distribution. For example, the first order time-varying L-moment $L_{i1}(t)$ simply represents the diurnal mean curve $X_i(t)$ aggregated into 10 minutes epoch (for $h=5$). The second order time-varying L-moment  $L_{i2}(t)$ captures a temporal change in variability and is similar to the diurnal standard deviation curve of physical activity considered by Varma and Watts \cite{varma2017daily}.

\begin{figure}[ht]
\centering
\includegraphics[width=.8\linewidth , height=.8\linewidth]{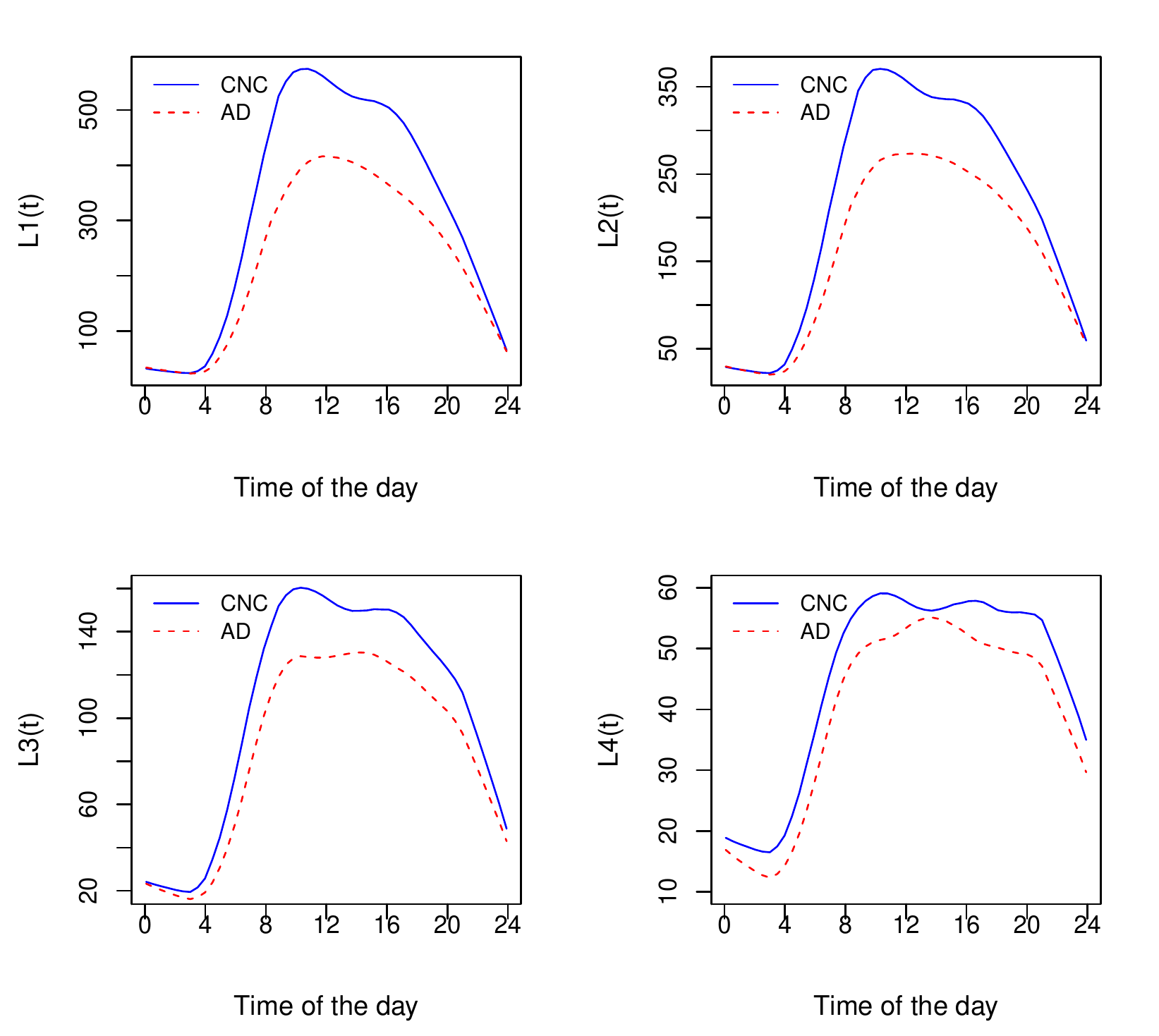}
\caption{The first four time-varying L-moments of daily physical activity averaged within CNC (blue) and AD (red) groups.}
\label{fig:fig3}
\end{figure}

Figure \ref{fig:fig3} displays the first four time-varying L-moments $L_r(t)$ of physical activity, averaged within CNC (blue) and AD (red) groups. Note that the first time-varying L-moments $L_1(t)$  exactly equal to the temporal diurnal curves from the top right panel of Figure \ref{fig:fig1c}. Subject-specific $r$-th order time-varying L-moment $L_{ir}(t)$ is related to the  time-by-distribution PA data object $Q_i(t,p)$ through its projection on Legendre polynomial basis $P_{r-1}(p)$ as follows
\begin{equation*}
    L_{ir}(t) = \int_0^1 Q_i(t,p)P_{r-1}(p)dp. 
\end{equation*} 
One can notice that mild AD have lower $L_1(t)$, $L_2(t)$, $L_3(t)$, and $L_4(t)$ moments compared to the CNC, particularly in the morning and somewhat in the afternoon.

We propose to use the time-varying subject-specific L-moments $L_{ir}(t)$ for modelling $Y_i$ using an additive SOFR model. If the shifted Legendre polynomials  $P_{\ell-1}(p)$ are used as the basis in $p$ for modelling the bivariate functional effect $\beta (t,p)$, the additive SOFR model (\ref{sofr4}) in terms of time-varying L-moments of PA provides an alternative representation of the bivariate SOFR model (\ref{sofr}) that is additionally interpretable from distributional point of view. We will refer to this approach as SOTDR-L. In particular, we have,
\begin{align}
g(\mu_i)&=\alpha+\*Z_i^T\bm\gamma + \int_{0}^{1} \int_{T}Q_i(t,p)\beta(t,p)dtdp \notag\\
&=\alpha+\*Z_i^T\bm\gamma + \int_{T}\sum_{k=1}^{K_0}\sum_{\ell=1}^{L_0}\theta_{k,\ell} B_{T,k}(t) \int_{0}^{1}Q_i(t,p)P_{\ell-1}(p) dt dp \notag 
\end{align}
\begin{align}
&= \alpha+\*Z_i^T\bm\gamma + \sum_{\ell=1}^{L_0}\int_{T}L_{i\ell}(t)\sum_{k=1}^{K_0}\theta_{k,\ell} B_{T,k}(t) dt \notag\\
&= \alpha+\*Z_i^T\bm\gamma + \sum_{\ell=1}^{L_0}\int_{T}L_{i\ell}(t)\beta_{\ell}^{*}(t) dt.\label{sofr4}
\end{align}
Here the functional regression coefficient $\beta^{*}_r(t)$ capture the diurnal time-varying effect of the $r$-th order time-varying L-moment on the response $Y_i$ at time $t$. Thus, we get an additive SOFR with time-varying L-moments. It is important to note that if $L_0 = 1$ we get exactly the SOFR model (\ref{sofr2}) that uses subject-specific temporal curves as predictors. Thus, SODTR-L model (\ref{sofr4}) strictly includes model (\ref{sofr2}). 
%CAN WE USE A SIMPLE ARGUMENT TO MAKE A STATEMENT FOR DDA?

\section{Application of SOTDR to modelling cognitive status and function in Alzheimer's Disease}
\label{appsec}
In this section, we apply SOTDR to model cognitive status and function in the Alzheimer's disease (AD) study and compare it to the three existing approaches including regression via scalar summaries, SOFR with temporal diurnal curves, SOFR with quantile functions.
The {\tt refund} package \citep{refund} in R \citep{Rsoft} is used for implementation of SOFR. First, we will model cognitive status (CNC vs AD) and the three cognitive scores of attention (ATTN), visual memory (VM), and executive function (EF) using the bivariate time-by-distribution data objects as illustrated in the Section \ref{method1}. Second, we alternatively use an additive SOFR with time-varying L-moments.

\subsection{SOTDR modelling of cognitive status}
We model cognitive status (CNC vs AD) using the SODTR model (\ref{sofr}) with an additive adjustment for age, sex and years of education. For comparison with existing approaches, we fit models (\ref{sofr1}), (\ref{sofr2}) and (\ref{sofr3}) using as predictors subject-specific average PA, diurnal PA curves, quantile PA functions, respectively. Ten minute diurnal PA curves have been calculated by aggregating minute-level data into 10 minutes epochs, resulting in subject-specific diurnal  PA curves $X_i(t)$ of length $144$. As mentioned earlier, since the participants of  the study were highly sedentary \citep{watts2016intra} such 10-minute  aggregation serves as pre-smoothing and  retains the key temporal patterns of PA. When we report predictive performance summaries such as area under the curve (AUC) of the receiver operating characteristic, we perform repeated cross-validation and report the average cross-validated AUC (cvAUC). The results of the analyses are presented in Table \ref{tab2comb}. 

\begin{table}[ht] \centering 
  \caption{The results of modelling cognitive status (CNC vs AD) and physical activity  using Model 1-4 with an adjustment for age, sex, and education. The standard deviation of the estimated coefficients for the scalar predictors are indicated in the parenthesis. Predictors: model 1-scalar average PA, model 2- diurnal PA curves, model 3 - quantile functions, model 4 - SOTDR with time-by-distribution data objects.} 
  \label{tab2comb} 
\begin{tabular}{@{\extracolsep{5pt}}lcccc} 
\\[-1.8ex]\hline 
\hline \\[-1.8ex] 
 & \multicolumn{4}{c}{\textit{Dependent variable : Cognitive Status (CNC vs AD)}} \\ 
\cline{2-5} 
\\[-1.8ex] & Model 1 & Model 2 & Model 3 & Model 4\\ 
\hline \\[-1.8ex] 
Intercept & 7.608$^{**}$ & 6.549$^{*}$ & 10.588$^{**}$& 12.368$^{***}$ \\ 
  & (3.567) & (3.615) & (4.139) & (4.591)\\ 
  & & & \\ 
 age & $-$0.051 & $-$0.040 & $-0.072^{*}$ &  $-0.089^{*}$\\ 
  & (0.038) & (0.039) & (0.043) & (0.047)\\ 
  & & & \\ 
 sex & 2.134$^{***}$ & 2.111$^{***}$ & 2.527$^{***}$ & 2.637$^{***}$ \\ 
  & (0.554) & (0.553) & (0.624) & (0.676)\\ 
  & & & \\ 
 education & $-$0.224$^{**}$ & $-$0.213$^{**}$ & $-$0.167$^{*}$ & $-$0.174$^{*}$\\ 
  & (0.091) & (0.091) & (0.092) & (0.095) \\ 
  & & & \\ 
 $\bar{X}_i$ & $-$0.005$^{***}$ & NA & NA & NA \\ 
  & (0.002) &  &  & \\ 
  & & & \\ 
  $X_i(t)$ & NA & $\hat{\beta}(t)^{**}$ & NA & NA \\ 
  &  &  &  &\\ 
  & & & \\ 
 $Q_i(p)$ & NA & NA & $\hat{\beta}(p)^{**}$ &NA \\ 
  &  &  &  & \\ 
  & & & \\ 
 $ Q_i(t,p)$ & NA & NA & NA &$\hat{\beta}(t,p)^{***}$ \\ 
  &  &  &  \\ 
  & & & \\ 
 
\hline \\[-1.8ex] 
Observations & 92 & 92 & 92 &92\\ 
cvAUC & 0.781 & 0.773 & 0.792 & 0.811\\ 
\hline 
\hline \\[-1.8ex] 
\textit{Note:}  & \multicolumn{3}{r}{$^{*}$p$<$0.1; $^{**}$p$<$0.05; $^{***}$p$<$0.01} \\ 
\end{tabular} 
\end{table} 

\begin{figure}[ht]
\begin{center}
\begin{tabular}{ll}
\includegraphics[width=.45\linewidth , height=.45\linewidth]{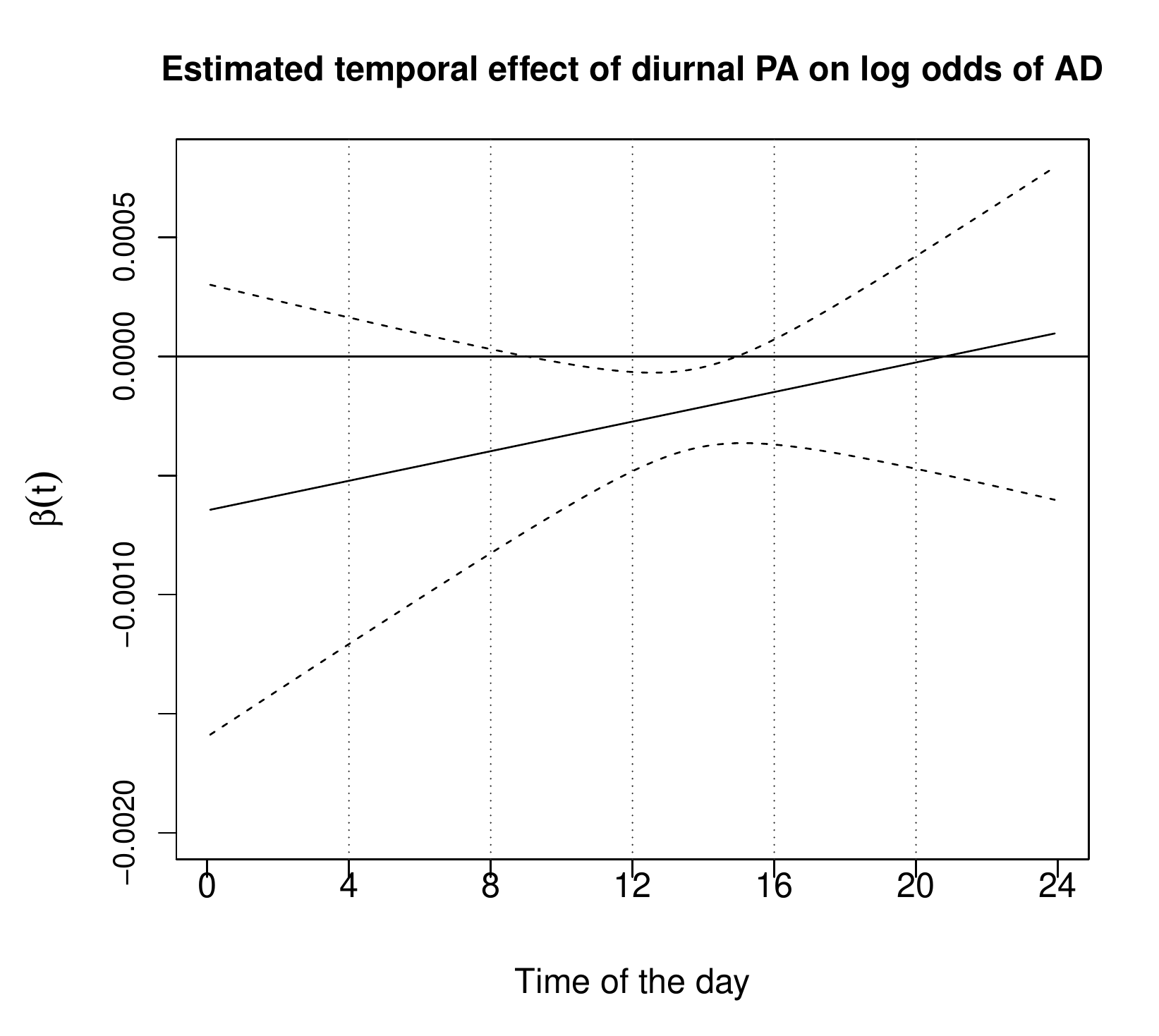} &
\includegraphics[width=.45\linewidth , height=.45\linewidth]{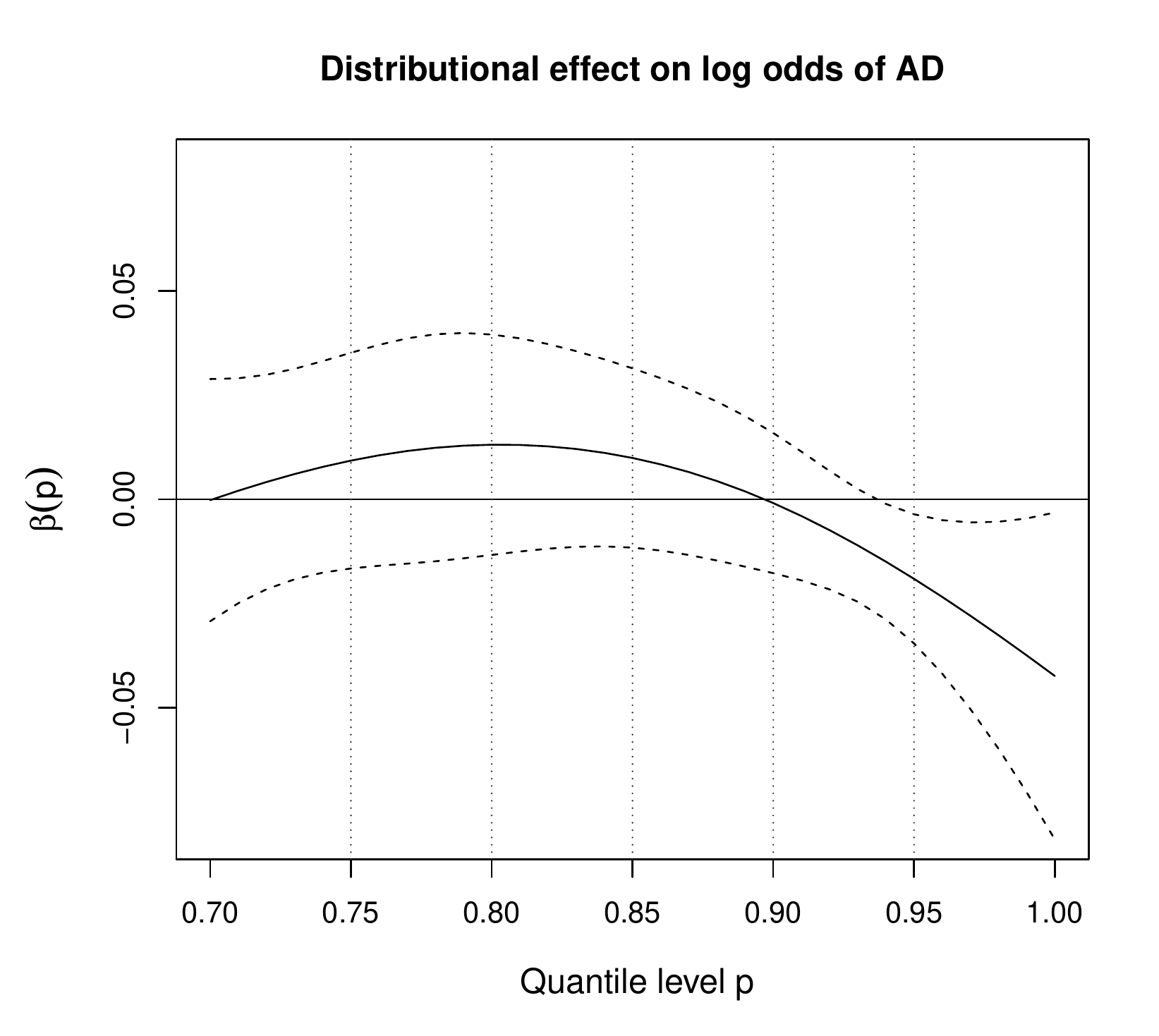}\\
\includegraphics[width=.52\linewidth , height=.52\linewidth]{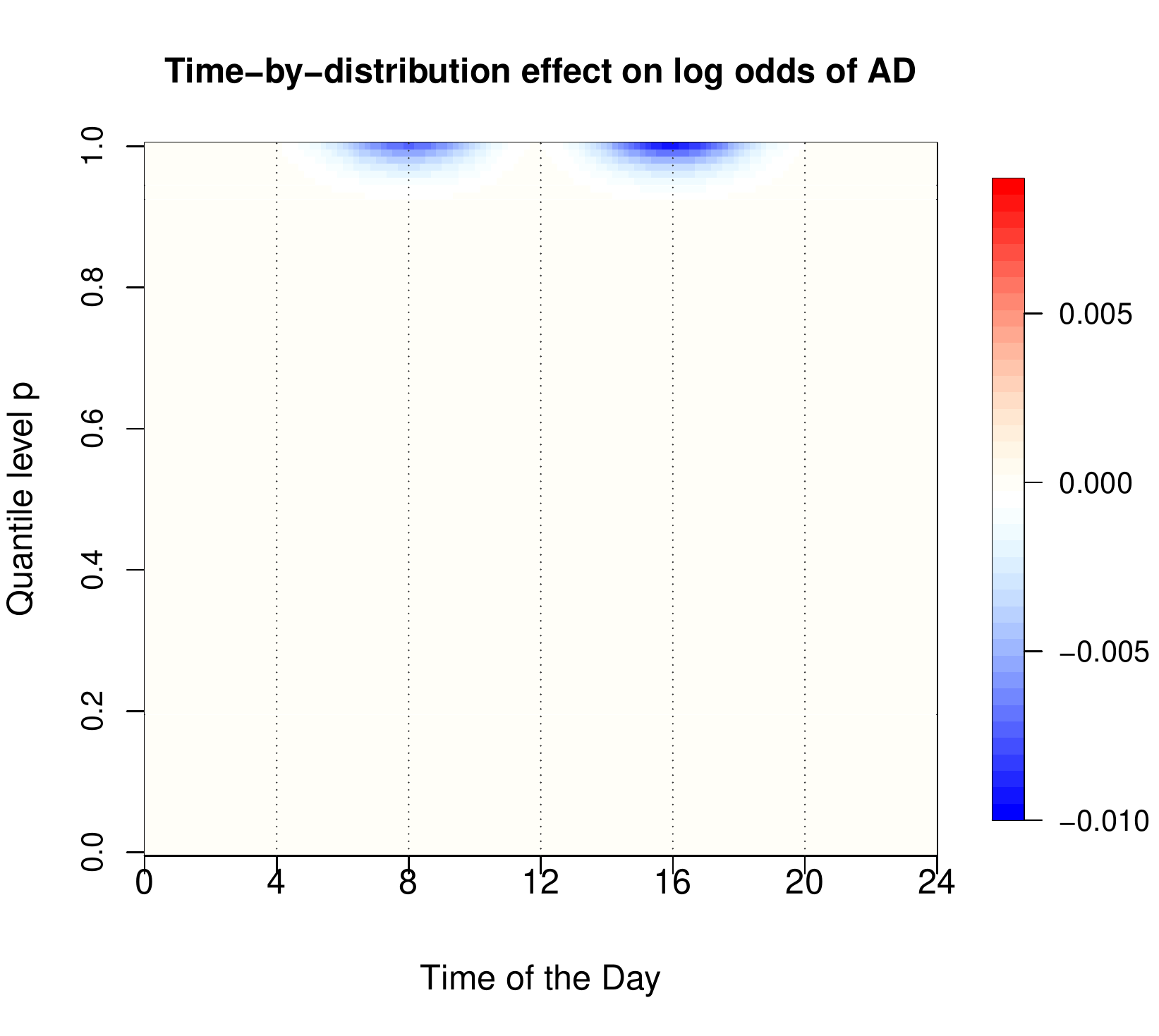}& \\
\end{tabular}
\end{center}
\caption{The estimated regression coefficients for Models 2-4. Estimated temporal effect $\beta(t)$ (top left. $t$ denoting time of the day). Estimated distributional effect $\beta(p)$ (top right, $p\in (0.7,1)$). Estimated bivariate effect $\beta(t,p)$ of time-by-distribution PA surface (bottomleft).}
\label{fig:fig3comb}
\end{figure}

Model 1 shows that higher subject-specific average PA is significantly
associated ($\alpha$= $0.05$) with a lower odds of AD. The cvAUC value of 0.781 illustrates a satisfactory discriminatory power of the model and is set as a benchmark for comparison with the other three models. The estimated functional regression coefficient $\beta(t)$ for Model 2 illustrating a diurnal effect of PA profile on log-odds of AD is displayed in the top left panel of Figure \ref{fig:fig3comb}. Model 2 finds that higher PA during morning hours ($\sim$ 10 am-3 p.m) is significantly associated ($\alpha=0.05$) with a lower odds of AD \citep{goldsmith2011penalized}. 

The average cvAUC of $0.773$ suggests that, although, the diurnal patterns of average PA offer additional temporal insights, they do not necessarily offer more discrimination between CNC and AD group compared to the use of simple average PA (Model 1, cvAUC = $0.781$). Model 3 finds the significance of subject-specific quantile functions of PA. The estimated functional regression coefficient $\beta(p)$ for Model 3 illustrating a distributional effect of PA on log-odds of AD is displayed ($\beta(p)$ not significant for $p<0.7$) in the top right panel of Figure \ref{fig:fig3comb} and shows that higher upper quantile levels ($p \in (0.90, 1)$) of PA are significantly associated with lower odds of AD \citep{goldsmith2011penalized}. Increased cvAUC of $0.792$ 
indicates higher discriminatory power of distributional encoding of PA (in particular, maximal PA) between CNC and AD compared to the average PA. 

The estimated bivariate functional effect $\beta(t,p)$ for Model 4 is shown in the bottomleft panel of Figure \ref{fig:fig3comb}. We used  $K_0=L_0=12$ cubic B-spline basis functions for modelling $\beta(t,p)$. Increased maximal capacity of PA during the morning ($\sim$ 7 a.m- 10 a.m) and in the afternoon ($\sim$ 3 p.m- 5 p.m) is found to be associated with lower odds of AD. An increased cvAUC of $0.811$ (around 3.8 $\%$ gain) illustrates additional discriminatory power of the time-by-distribution PA data objects, while simultaneously capturing temporally local distributional effects of the PA on log-odds of AD.

\subsection{SOTDR-L modelling of cognitive status}
Next, we illustrate an application of SOTDR-L that uses diurnal time-varying L-moments for modelling cognitive status (CNC vs AD) outcome. For interpretability, we use the first four diurnal L-moments profile $L_{ik}(t)$ ($L_0=4$) as functional predictors and adjust for age, sex and years of education. Since we have a relatively small sample size (n=92), we follow a penalized SOFR approach
\begin{figure}[ht]
\begin{center}
\includegraphics[width=.45\linewidth , height=.45\linewidth]{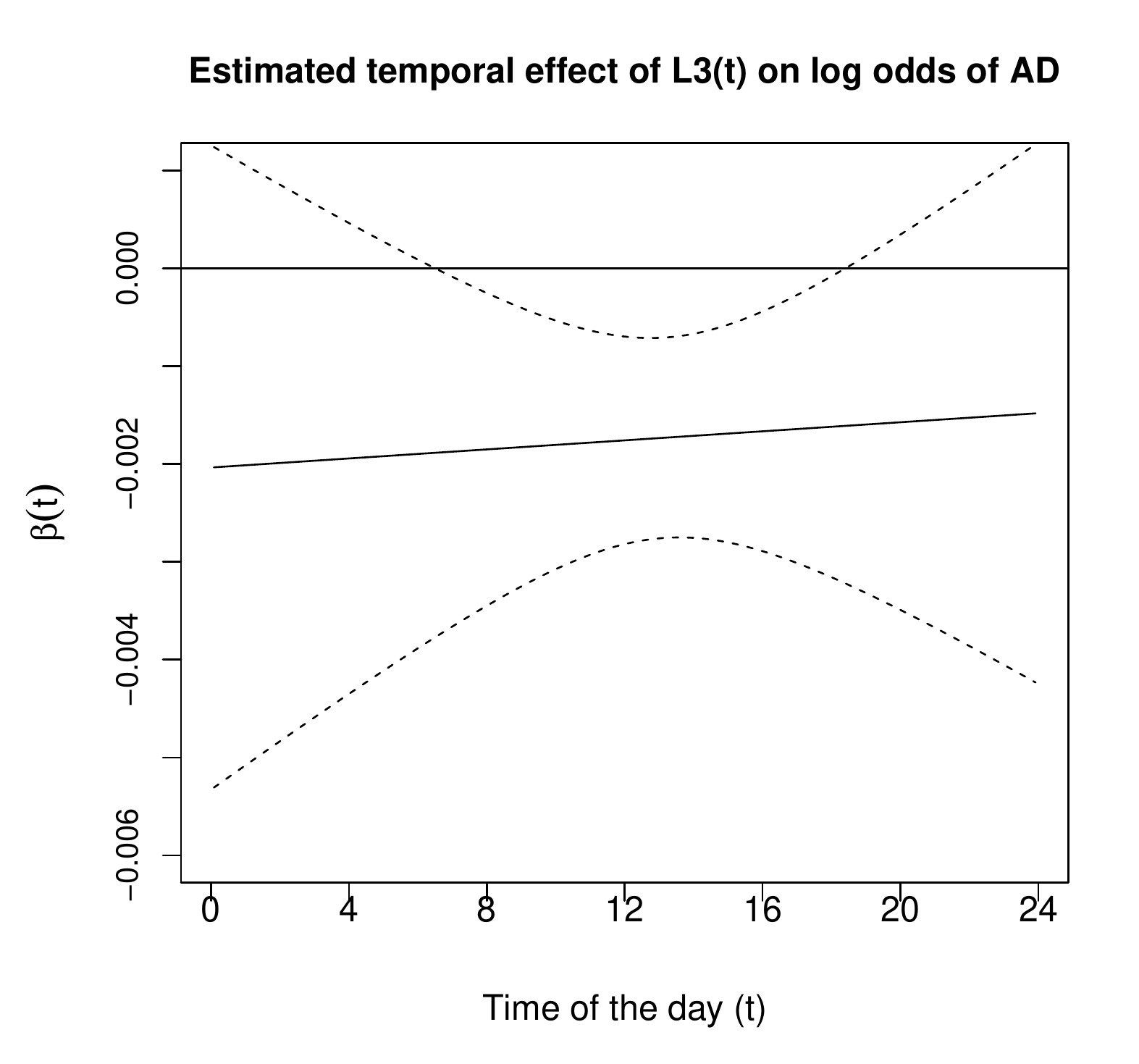} 
\end{center}
\caption{Estimated diurnal effect $\beta(t)$ of $L_{i3}(t)$ of PA on log odds of AD.}
\label{fig:fig7}
\end{figure}
to select the L-moments $L_{ik}(t)$-s, which are most informative. 
In particular, we re-express the SOFR model (\ref{sofr4}) in terms of functional principal component scores of $L_{ik}(t)$ following a functional principal component regression approach \citep{reiss2007functional,kong2016classical}, 
\begin{eqnarray}
E(Y_i|\{L_{ir}(t)\}_{r=1}^4)=\mu_i, \hspace{2 mm}
g(\mu_i)=\alpha+\*Z_i^T\bm\gamma + \sum_{r=1}^{4}\sum_{s=1}^{n_r}\xi_{irs}\beta_{r,s}\label{sofrLre}.
\end{eqnarray}
Here $\xi_{irs}=\int_{T}L_{ir}(t)\psi_{s}(t)$ is the projection of the diurnal L-moment $L_{ir}(t)$ on the eigenbasis $\psi_{s}(t)$ and $\beta_r(t)$ is modelled as $\beta_r(t)=\sum_{s=1}^{n_r}\beta_{r,s}\psi_{s}(t)$. We use the group exponential Lasso (GEL) penalty \citep{breheny2015group} on the basis coefficients $\{\beta_{r,s}\}_{s=1}^{n_r}$ to perform variable selection in order to identify informative time-varying L-moments $L_{ik}(t)$. GEL is a bi-level selection penalty and enjoys the added flexibility of forcing some of the coefficients within a particular group to be zero, thus effectively reducing the number of parameters, which is especially useful in our scenario due to the very low sample size. The proposed variable selection approach selects the 3rd order time-varying L-moments $L_{i3}(t)$ to be most informative i.e, most discriminating between the two groups (CNC and AD) while adjusting for age, sex and years of education. The \texttt{grpreg} package \citep{bre2015} in R is used for implementing the variable selection method using GEL.
The estimated diurnal effect of $L_{i3}(t)$ is shown in Figure \ref{fig:fig7}. We observe that an increase in the value of third order L-moment of physical activity,  during the window (8 a.m- 6 p.m) is associated with a lower odds of AD. The third order L-moment $L_{i3}(t)$ is related to L-skewness of the PA and its significance is therefore very interesting from a clinical perspective.
We also perform repeated cross-validation using $L_{i3}(t)$ as predictor in a SOFR model while adjusting for age, sex, and years of education. An increased cvAUC of $0.802$ (around 2.7 $\%$ gain) illustrates satisfactory discriminatory power of the proposed metric offering both distributional and temporal encoding of physical activity. Likely, because of restricting the number of L-moments and the use of GEL, the temporal findings of SOTDR-L differ from temporal findings of SOTDR. While SOTDR  highlights activity in the upper quantile levels during 6-10am and 2-6pm time periods, SOTDR-L highlights the third order L-moment of activity during mid-day hours that are similar to those from SOFR on temporal diurnal curves. Chosen third order time-varying L-moments in SOTDR-L also seems to result in an increase in cvAUC compared to SOFR that uses temporal diurnal curves (that are equivalent to the first order time-varying L-moments).  %%This difference between times choosen worth exploring. Most likely is due to the small sample sizes.

\subsection{Modelling attention}
In this section, we apply SOTDR to the cognitive score of attention (ATTN) and the results are compared with those from regression with subject-specific average PA, FDA using diurnal PA curves, DDA using quantile functions. Adjusted R-squared, defined as the adjusted proportion of variance explained, where original variance and residual variance are both estimated using unbiased estimators \cite{wood2017generalized}, is used in Models 2-4 for the evaluation of predictive performance. 

\begin{table}[ht] \centering 
  \caption{The results of modelling attention score and physical activity  using Model 1-4 with an adjustment for age, sex, and education. The standard deviation of the estimated coefficients for the scalar predictors are indicated in the parenthesis. Predictors: model 1-scalar average PA, model 2- diurnal PA curves, model 3 - quantile functions, model 4 - SOTDR with time-by-distribution data objects. All models are adjusted for age, sex, years of education.} 
  \label{tab3comb} 
\begin{tabular}{@{\extracolsep{5pt}}lcccc} 
\\[-1.8ex]\hline 
\hline \\[-1.8ex] 
 & \multicolumn{4}{c}{\textit{Dependent variable : ATTN score}} \\ 
\cline{2-5} 
\\[-1.8ex] & Model 1 & Model 2 & Model 3 & Model 4\\ 
\hline \\[-1.8ex] 
Intercept & $-$1.423 & $-$1.157 & $-2.045^{**}$ & -3.696$^{***}$ \\ 
  & (0.929) & (0.960) & (0.927) & (0.988)\\ 
  & & & \\ 
 age & 0.002 & -0.001 & 0.006 &  $0.021^{*}$\\ 
  & (0.011) & (0.011) & (0.010) & (0.011)\\ 
  & & & \\ 
 sex & $-$0.354$^{**}$ & $-$0.349$^{**}$ & $-$0.443$^{***}$ & $-0.476^{***}$ \\ 
  & (0.150) & (0.150) & (0.150) & (0.134)\\ 
  & & & \\ 
 education & 0.083$^{***}$ & 0.080$^{***}$ & 0.072$^{***}$ & $0.069^{***}$\\ 
  & (0.023) & (0.023) & (0.023) & (0.021) \\ 
  & & & \\ 
 $\bar{X}_{i}$ & 0.0005 & NA & NA & NA \\ 
  & (0.0005) &  &  & \\ 
  & & & \\ 
 $X_i(t)$ & NA & $\hat{\beta}(t)$ & NA & NA \\ 
  &  &  &  &\\ 
  & & & \\ 
  $Q_i(p)$ & NA & NA & $\hat{\beta}(p)^{**}$ &NA \\ 
  &  &  &  & \\ 
  & & & \\ 
  $Q_{i}(t,p)$ & NA & NA & NA &$\hat{\beta}(t,p)^{***}$ \\ 
  &  &  &  \\ 
  & & & \\ 
 
\hline \\[-1.8ex] 
Observations & 92 & 92 & 92 &92\\ 
Adjusted R$^{2}$ & 0.161 & 0.163 & 0.218 & 0.378\\ 
\hline 
\hline \\[-1.8ex] 
\textit{Note:}  & \multicolumn{3}{r}{$^{*}$p$<$0.1; $^{**}$p$<$0.05; $^{***}$p$<$0.01} \\ 
\end{tabular} 
\end{table} 

\begin{figure}[ht]
\begin{center}
\begin{tabular}{ll}
\includegraphics[width=.45\linewidth , height=.45\linewidth]{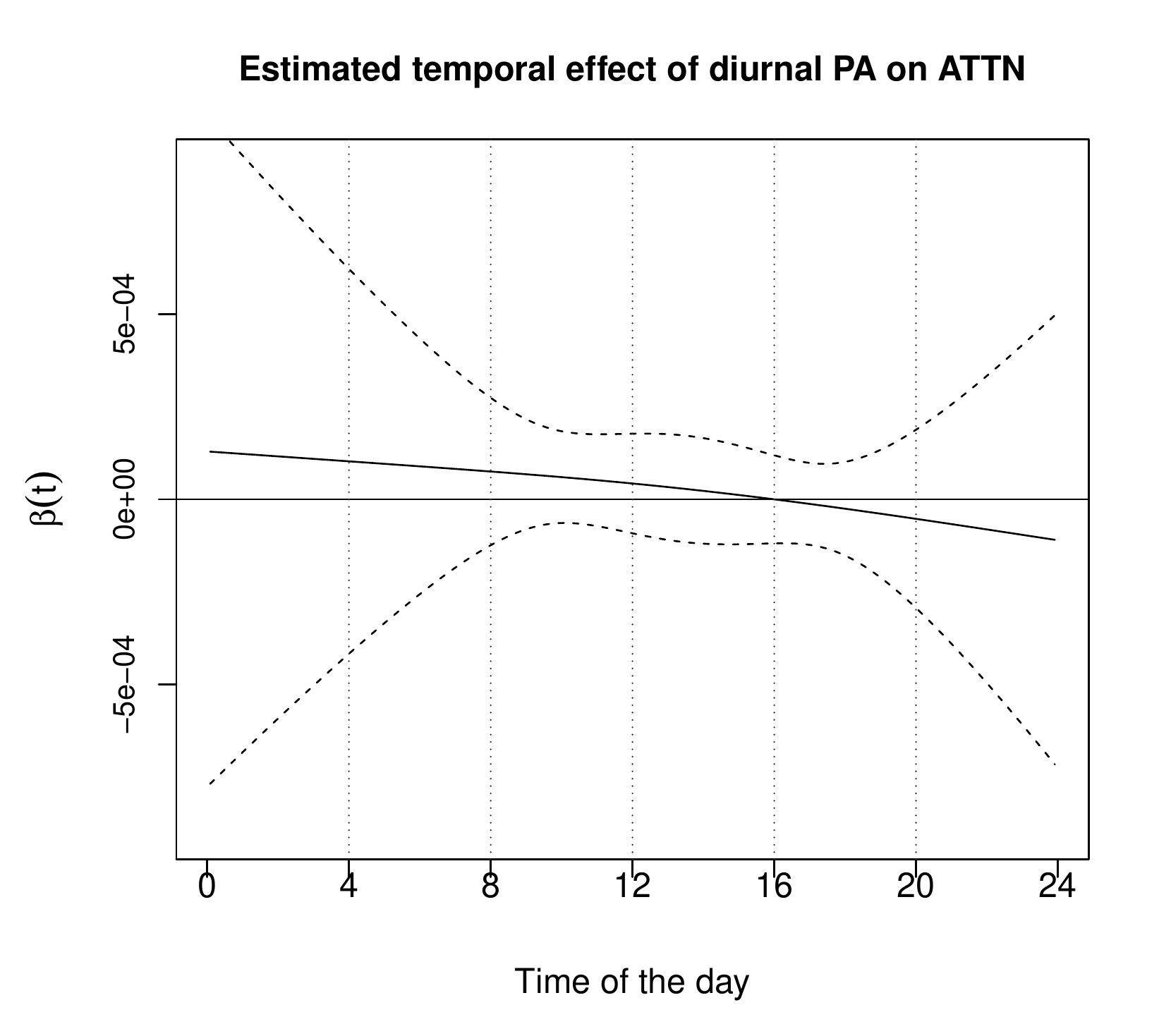} &
\includegraphics[width=.45\linewidth , height=.45\linewidth]{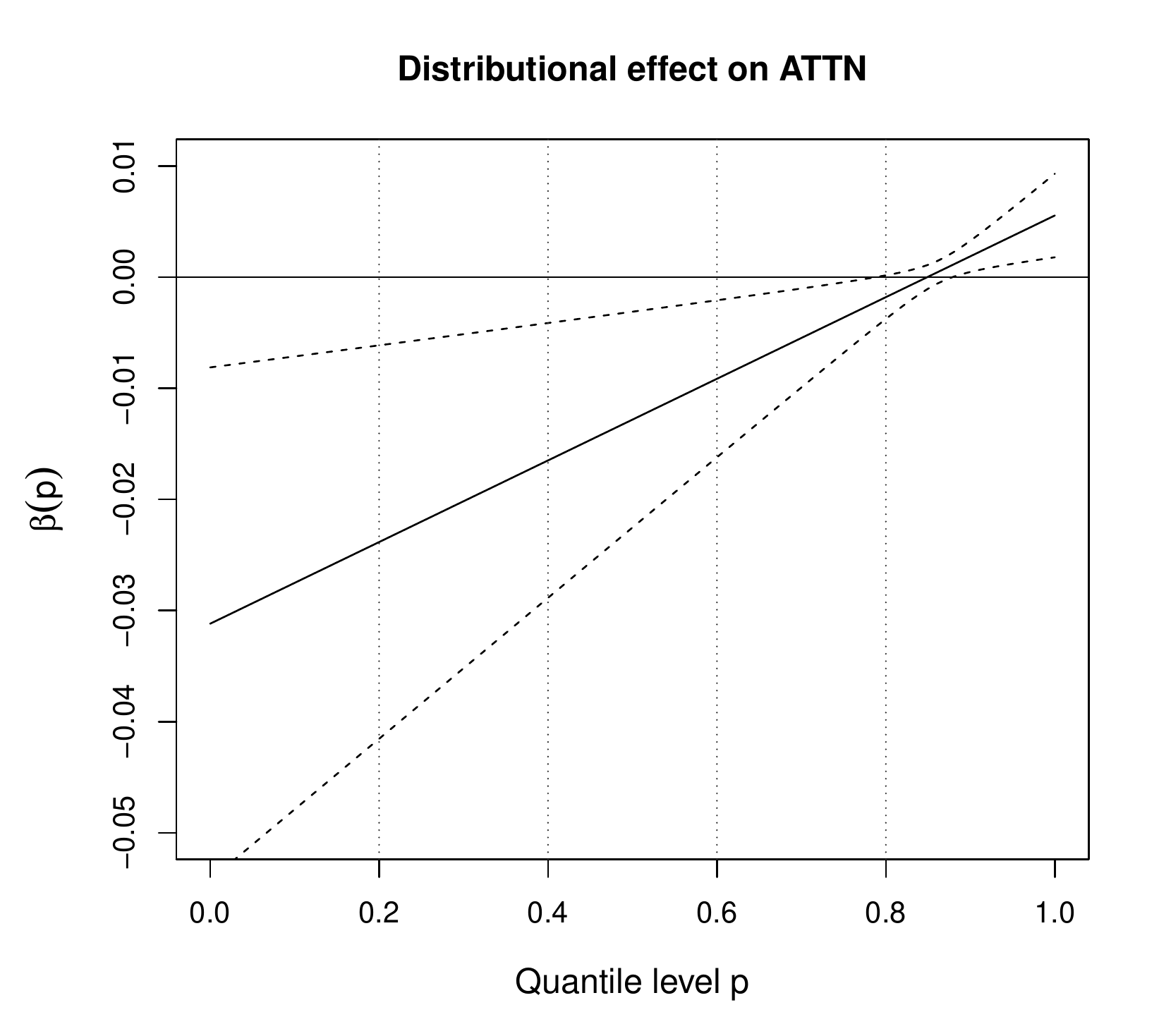}\\
\includegraphics[width=.5\linewidth , height=.5\linewidth]{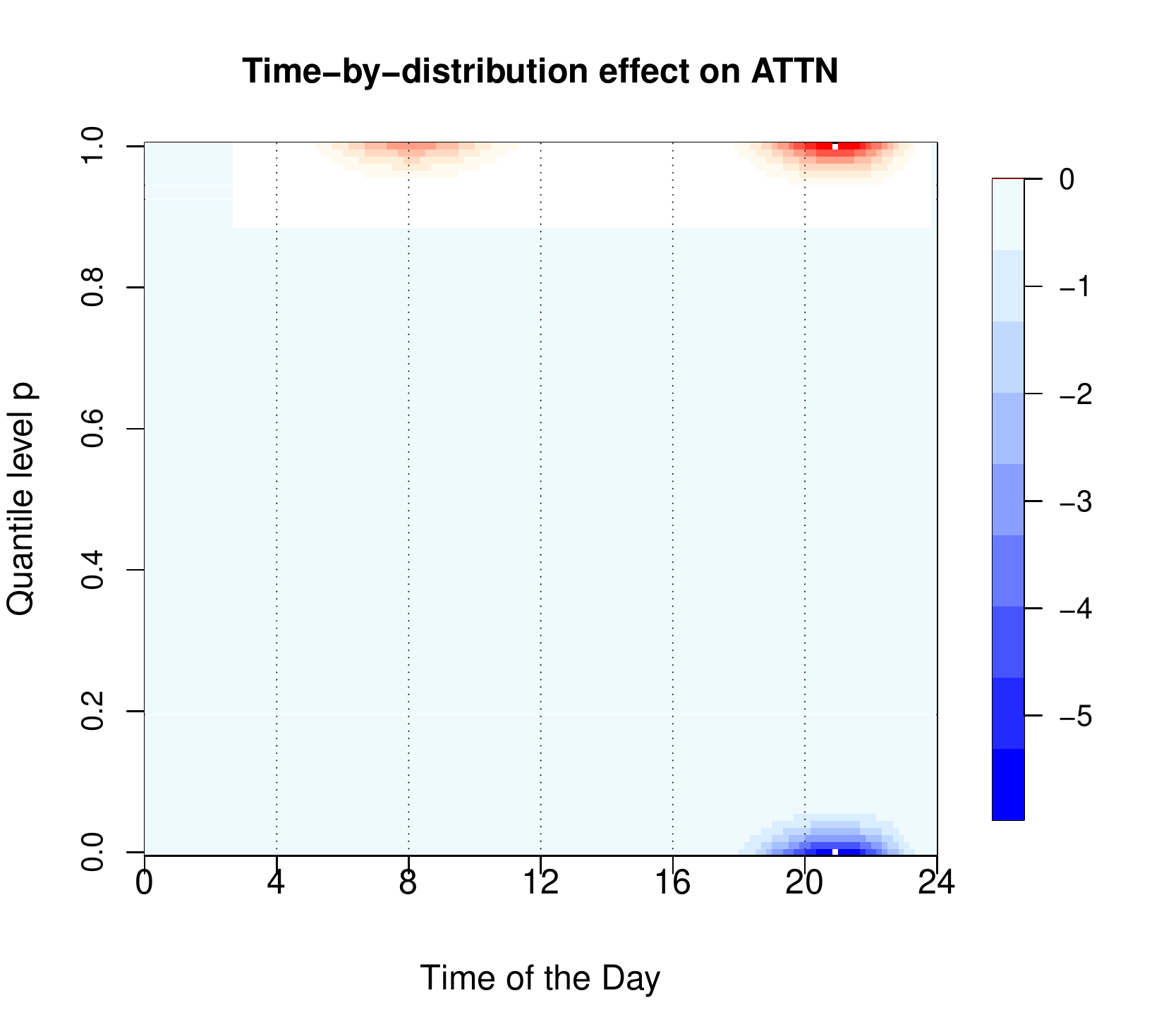}&
\includegraphics[width=.5\linewidth , height=.5\linewidth]{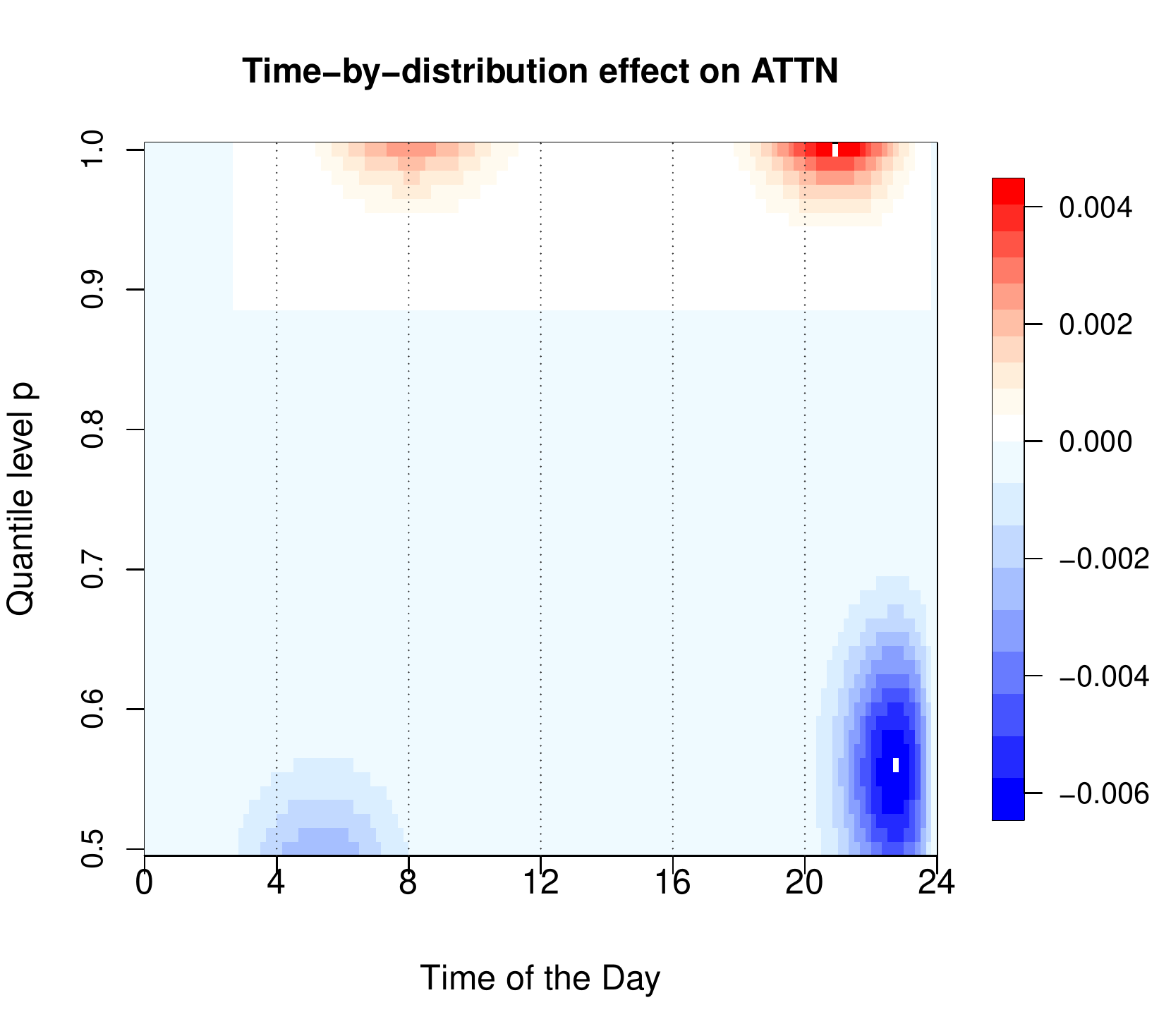}
\\
\end{tabular}
\end{center}
\caption{The estimated effects of the different PA metrics (Model 2-4) on ATTN score. Estimated temporal effect  (solid line) $\beta(t)$ (top left). Estimated distributional effect $\beta(p)$ (top right). Estimated bivariate effect $\beta(t,p)$ of time-by-distribution PA surface (bottomleft). The same plot with $p$ restricted to the distributional domain $(0.5,1)$ (bottomright).}
\label{fig:fig4comb}
\end{figure}

Table \ref{tab3comb} presents the result of the analyses from these four modelling approaches. The association between average PA and attention is not found to be significant at $\alpha=0.05$ level.  Adjusted R-squared of the model is reported to be $0.161$ and is set as the benchmark for comparison with the other approaches. Although, the diurnal curves of PA were not found to be significant ($\alpha=0.05$ level), the estimated quantile-function effect is significant. The estimated regression coefficient $\beta(p)$ is shown in Figure \ref{fig:fig4comb} (top right panel). It shows that $\beta(p)$ creates a contrast between a higher quantile levels ($p > 0.8$) and lower quantile levels ($p < 0.8$). Specifically, an increase in higher quantile of PA is found to be associated with higher performance on attention. Although, one needs to be cautious in interpreting the results as subject- specific quantiles of PA are mostly zero below the quantile level $p<0.5$ as illustrated in Figure \ref{fig:fig1c}. A $35\%$ increase in the adjusted R-squared is observed using DDA with subject-specific quantile functions of PA compared to the benchmark model.

The estimated bivariate coefficient $\beta(t,p)$, capturing the TD effect on attention is displayed in Figure \ref{fig:fig4comb} (bottomleft panel). Increased maximal capacity of PA during the morning ($\sim$ 7 a.m - 10 a.m) and in the evening ($\sim$ 8 p.m - 10 p.m) is found to be associated with higher attention score after adjusting for age, sex and years of education. Importantly, when quantile levels are constrained to be above 0.5 (re-estimated $\beta(t,p)$ is shown in the bottomright), there are two contrasts between upper quantile levels ($p > 0.9$) and lower quantile levels ($0.5 < p < 0.7$) which are not time-aligned and actually capture quantile contrast between adjacent time periods. The morning TD effect can be interpreted as lower level quantile activity centered around 4-6 a.m are negatively associated and higher level quantile activity centered around 7-9 a.m are positively associated with attention. The evening TD effect can be interpreted higher level quantile activity centered around 8-10 p.m are positively associated and lower level quantile activity centered around 10 p.m-12 a.m are negatively associated with with attention.  Adjusted R-squared of SOTDR model using the time-by-distribution PA surface is reported to be $0.378$, giving a $135\%$ gain from the benchmark model using average physical activity, demonstrating very strong time-by-distribution effect, compared to the non-significant average and diurnal effect and significant distributional effect.
The results from the similar SOTDR analysis of verbal memory (VM) and executive function (EF) are presented in the Supplementary Tables A1, A2 and Supplementary Figures B1, B2 of the Appendix. For both outcomes, we observed significant improvements in  adjusted R-squared.

\subsection{SOTDR-based scalar biomarkers}

\begin{figure}[ht]
\begin{center}
\begin{tabular}{ll}
\includegraphics[width=.5\linewidth , height=.5\linewidth]{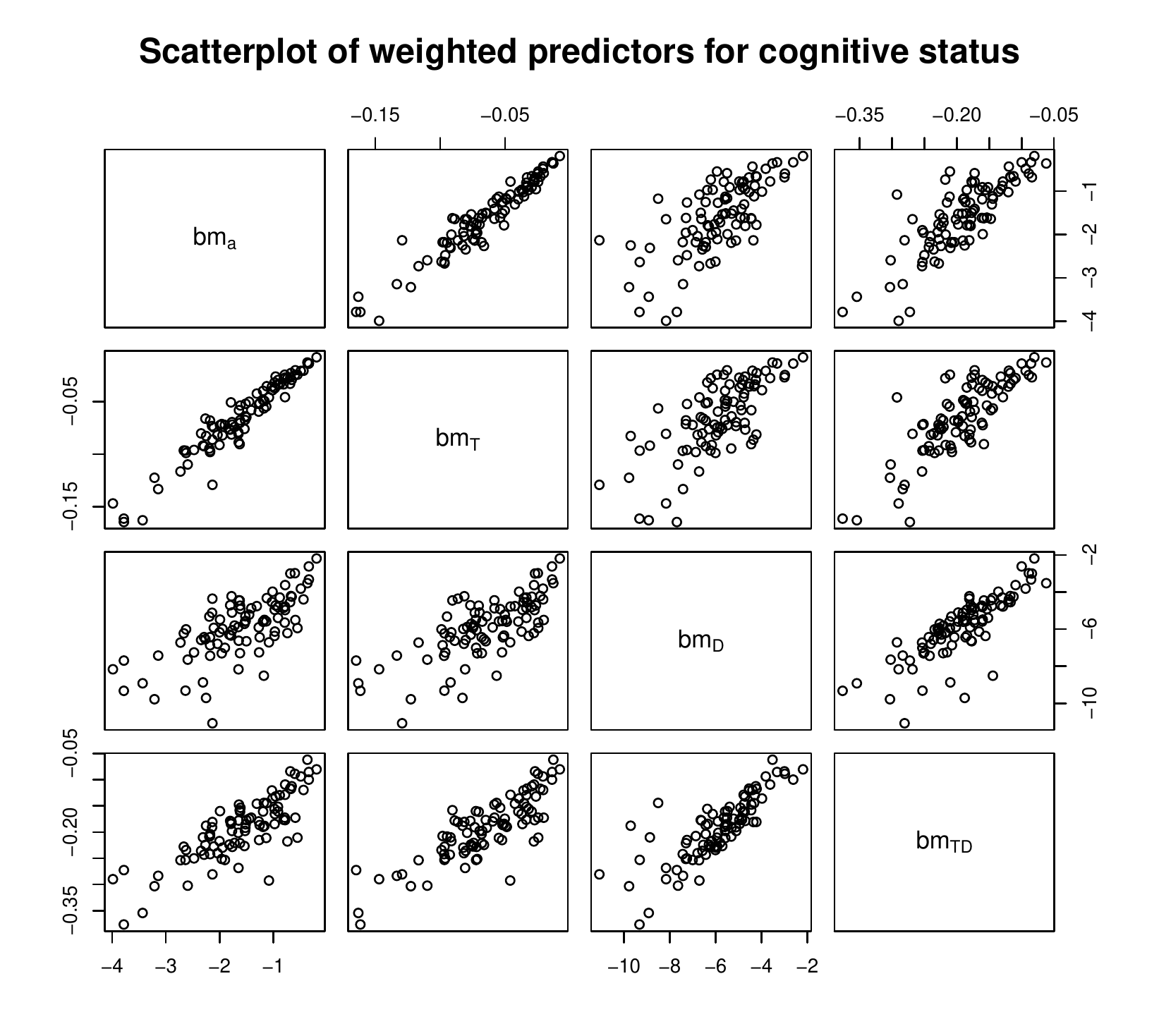} &
\includegraphics[width=.5\linewidth , height=.5\linewidth]{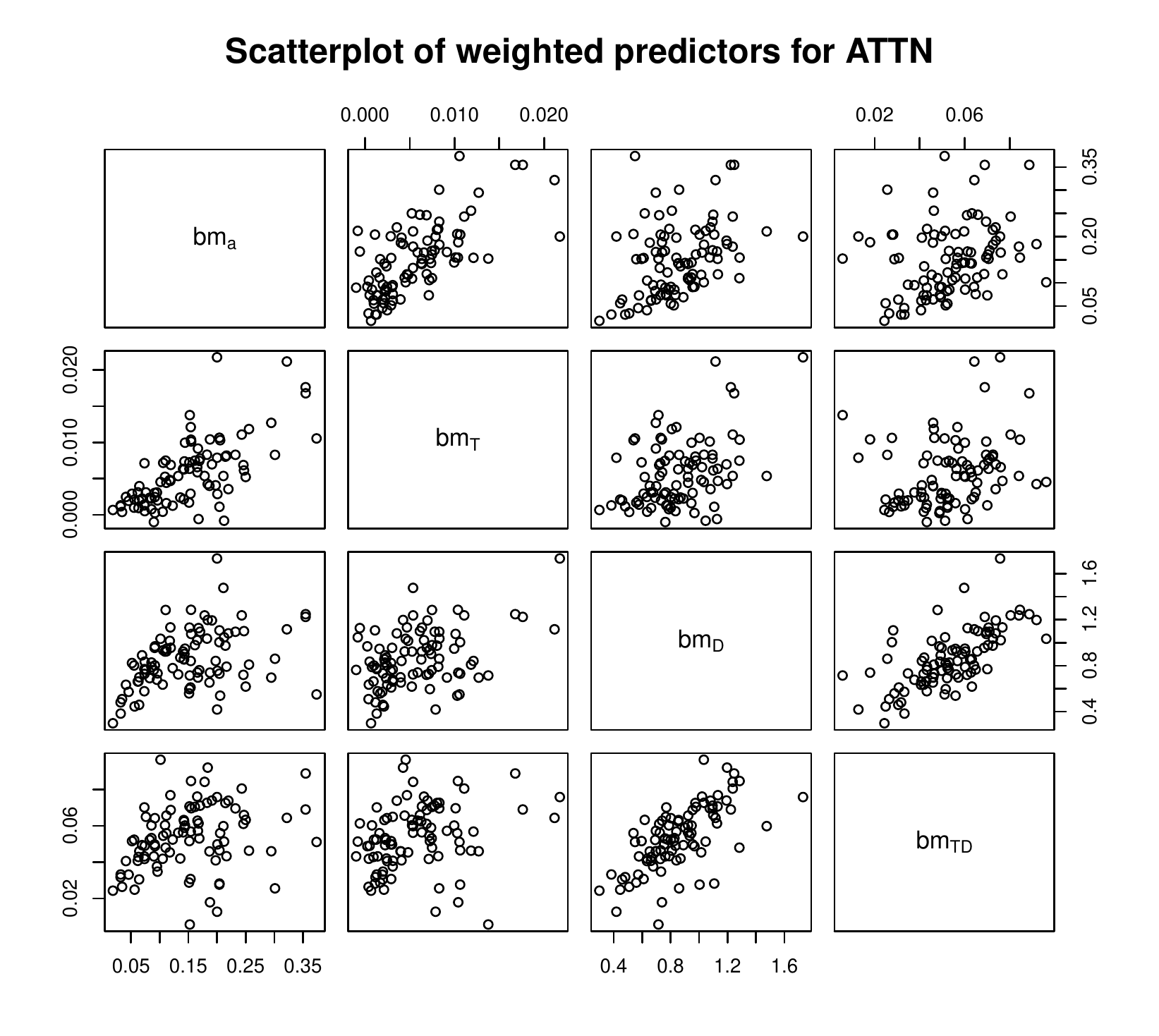}\\
\end{tabular}
\end{center}
\caption{scatterplots of the estimated weighted scores corresponding to the predictors average PA, diurnal PA curve, PA quantile function and time-by-distribution PA metric respectively for cognitive status (left) and ATTN (right).}
\label{fig:fig5}
\end{figure}

Estimated SOTDR can be used to create a simpler to use and interpret scalar biomarkers. For example, based on the previously fitted models for an outcome of interest, one can calculate following SOTDR biomarkers $bm_{TD,i} =\int_{0}^{1}  \int_{T}Q_i(t,p)\hat{\beta}(t,p)dtdp$ and compare them with the models based on the average PA, diurnal curves of PA, and quantile functions of PA: $bm_{a,i} = \bar{X}_i\hat{\beta}$,  $bm_{T,i} =\int_{T} X_i(t)\hat{\beta}(t)dt$,  $bm_{D,i} = \int_{0}^{1} Q_i(p)\hat{\beta}(p)dp$.  
Figure \ref{fig:fig5} displays the scatterplot matrix for all four types of biomarkers to discriminate either cognitive status (left) and attention score (right). Although, they are mostly positively correlated, the large amount of spread indicates that they likely capture somewhat different aspects of PA.

\section{Discussion}\label{sec6}
In this paper, we have used subject-specific time-by-distribution data objects to capture and model temporally local distributional information in wearable data. We then developed a scalar on time-by-distribution regression that handles TD objects as predictors. We have also provided an alternative and parsimonious representation of the time-by-distribution objects in terms of time-varying L-moments, robust rank-based analogs of traditional moments. This representation allowed us to illustrate that SOTDR generalizes SOFR.

Our approach revealed novel insights in the associations between  distributional and diurnal aspects of physical activity and various domains of cognitive function and Alzheimer’s disease status. The time-by-distribution representation provided better discrimination between the CNC and AD participants. Our results revealed strong associations between temporally local distributional aspects of PA across the day and clinical cognitive scales impacted in early AD, especially, attention. These results highlight the potential value of designing and testing physical activity interventions targeting specific time of the day, in the early stages of AD. As there may be times of the day when cognitively impaired individuals are most alert \citep{musiek2015sleep,volicer2001sundowning}, it might be specifically suited for individual specific PA interventions. 
Note that, although we have not established a causal direction here, it could also be that people with AD have poor sleep, so are less active in the morning compared to cognitively normal controls. The maximal capacity of physical activity represents reserve of a individual and our study has revealed strong and significant associations between cognitive performance and maximal PA levels, indicating changes in the reserve of a person might be sensitive to specific disease pathology and cognitive decline.

This paper opens interesting research questions on how to efficiently capture information with TD data objects. In our approach we encoded distributional information via quantile functions, the use of other distributional representation such as CDF or hazard function could be explored in future work. In our application, the window length $h$ for calculating $Q_i(t,p)$ and $L_i(t)$ was chosen to be consistent with the window size for diurnal curves. However, in other applications, an adaptive procedure of the choice of optimal window size $h$ may be developed. Time registration or time-warping is often a desirable pre-processing step to make sure the amplitude and phase variations in functional data are properly separated \citep{dryden2016statistical, wrobel2019registration,marron2015functional}. This is, especially, important for wearable data which often driven by subject-specific schedules and time preferences. Thus, pre-registration of TD objects is another exciting area of future research. We have focused on a linear effect of the TD data objects in this paper due to it's simplicity, interpretability and connection with summary level modelling approaches. Accounting for circular nature of the data may be another interesting direction.  Future applications might benefit from considering nonlinear effects of the  TD objects and this could be done via nonlinear extensions scalar-on-function regression models \citep{reiss2017methods}. Another interesting area of research would be to extend and apply the proposed method for modelling longitudinal data that at each visit generate distribution. For example, this could offer novel insights into the distributional changes in physical activity at different ages across the lifespan. \citep{varma2017re}.

%\backmatter

%\section*{Acknowledgments}
%This is acknowledgment text.\cite{Kenamond2013} Provide text here. This is acknowledgment text. Provide text here. This is acknowledgment text. Provide text here. This is acknowledgment text. Provide text here. This is acknowledgment text. Provide text here. This is acknowledgment text. Provide text here. This is acknowledgment text. Provide text here. This is acknowledgment text. Provide text here. This is acknowledgment text. Provide text here. 

%\subsection*{Author contributions}

%This is an author contribution text. This is an author contribution text. This is an author contribution text. This is an author contribution text. This is an author contribution text. 

%\subsection*{Financial disclosure}

%None reported.
\subsection*{R Vignette}
Illustration of the proposed framework via R \citep{Rsoft}, along with the dataset analyzed, is available online with this article and on Github at \url{https://github.com/rahulfrodo/SOTDR}.

\subsection*{Conflict of interest}

The authors declare no potential conflict of interests.

\section*{Supporting information}

Supplementary Tables A1, A2 and Supplementary Figures B1, B2 are available as supporting information as part of the online article.

%\noindent
%\textbf{Figure S1.}
%{500{\uns}hPa geopotential anomalies for GC2C calculated against %the ERA Interim reanalysis. The period is 1989--2008.}

%\noindent
%\textbf{Figure S2.}
%{The SST anomalies for GC2C calculated against the observations %(OIsst).}

%\nocite{*}% Show all bib entries - both cited and uncited; comment this line to view only cited bib entries;
\newpage
\bibliography{main}%
%\section*{Author Biography}

%\begin{biography}{\includegraphics[width=66pt,height=86pt,draft]{empty}}{\textbf{Author Name.} This is sample author biography text this is sample author biography text this is sample author biography text this is sample author biography text this is sample author biography text this is sample author biography text this is sample author biography text this is sample author biography text this is sample author biography text this is sample author biography text this is sample author biography text this is sample author biography text this is sample author biography text this is sample author biography text this is sample author biography text this is sample author biography text this is sample author biography text this is sample author biography text this is sample author biography text this is sample author biography text this is sample author biography text.}
%\end{biography}
\appendix
%table for VM
\section{Supplementary Tables}
\begin{table}[H] \centering 
  \caption{Results from modelling cognitive score of VM on age, sex, education and physical activity metrics using Model 1-4. The standard deviation of the estimated coefficients for the scalar predictors are indicated in the parenthesis. Model 1: summary level modelling using average PA, Model 2: Temporal modelling using diurnal PA, Model 3: Distributional modelling using PA quantile function, Model 4: Joint modelling using PA time-by-distribution bivariate surface.} 
  \label{tab3combvm} 
\begin{tabular}{@{\extracolsep{5pt}}lcccc} 
\\[-1.8ex]\hline 
\hline \\[-1.8ex] 
 & \multicolumn{4}{c}{\textit{Dependent variable : VM score}} \\ 
\cline{2-5} 
\\[-1.8ex] & Model 1 & Model 2 & Model 3 & Model 4\\ 
\hline \\[-1.8ex] 
Intercept & $-$2.329 & $-$1.561 & $-3.635^{*}$ & -3.786$^{*}$ \\ 
  & (1.950) & (2.006) & (1.948) & (2.061)\\ 
  & & & \\ 
 age &$-$0.009 & $-$0.017 & 0.001 &  $-0.0001$\\ 
  & (0.023) & (0.023) & (0.022) & (0.023)\\ 
  & & & \\ 
 Sex & $-$1.355$^{***}$ & $-$1.338$^{***}$ & $-$1.533$^{***}$ & $-1.35^{***}$ \\ 
  & (0.315) & (0.313) & (0.312) & (0.306)\\ 
  & & & \\ 
 education & 0.164$^{***}$ &0.156$^{***}$ & 0.142$^{***}$ & $0.126^{***}$\\ 
  & (0.049) & (0.049) & (0.048) & (0.047) \\ 
  & & & \\ 
 $\bar{X}_{i}$ & 0.003$^{***}$ & NA & NA & NA \\ 
  & (0.001) &  &  & \\ 
  & & & \\ 
  $X_i(t)$ & NA &  $\hat{\beta}(t)^{***}$ & NA & NA \\ 
  &  &  &  &\\ 
  & & & \\ 
 $Q_i(p)$ & NA & NA & $\hat{\beta}(p)^{***}$ &NA \\ 
  &  &  &  & \\ 
  & & & \\ 
 $Q_{i}(t,p)$ & NA & NA & NA &$\hat{\beta}(t,p)^{***}$ \\ 
  &  &  &  \\ 
  & & & \\ 
 
\hline \\[-1.8ex] 
Observations & 92 & 92 & 92 &92\\ 
Adjusted R$^{2}$ & 0.331 & 0.338 & 0.375 & 0.413\\ 
\hline 
\hline \\[-1.8ex] 
\textit{Note:}  & \multicolumn{3}{r}{$^{*}$p$<$0.1; $^{**}$p$<$0.05; $^{***}$p$<$0.01} \\ 
\end{tabular} 
\end{table}

%table for EF
\begin{table}[H] \centering 
  \caption{Results from modelling cognitive score of EF on age, sex, education and physical activity metrics using Model 1-4. The standard deviation of the estimated coefficients for the scalar predictors are indicated in the parenthesis. Model 1: summary level modelling using average PA, Model 2: Temporal modelling using diurnal PA, Model 3: Distributional modelling using PA quantile function, Model 4: Joint modelling using PA time-by-distribution bivariate surface.} 
  \label{tab3combef} 
\begin{tabular}{@{\extracolsep{5pt}}lcccc} 
\\[-1.8ex]\hline 
\hline \\[-1.8ex] 
 & \multicolumn{4}{c}{\textit{Dependent variable : EF score}} \\ 
\cline{2-5} 
\\[-1.8ex] & Model 1 & Model 2 & Model 3 & Model 4\\ 
\hline \\[-1.8ex] 
Intercept & $-$2.479 & $-$1.944 & $-3.070^{**}$ & $-$3.760$^{**}$ \\ 
  & (1.492) & (1.539) & (1.531) & (1.612)\\ 
  & & & \\ 
 age &$-$0.0002 &$-$0.006 &$0.004$ &  0.009\\ 
  & (0.017) & (0.018) & (0.017) & (0.018)\\ 
  & & & \\ 
 Sex & $-$1.063$^{***}$ & $-$1.051$^{***}$ & $-$1.141$^{***}$ & $-1.094^{***}$ \\ 
  & (0.241) & (0.240) & (0.245) & (0.230)\\ 
  & & & \\ 
 education & 0.141$^{***}$ &0.136$^{***}$ & 0.132$^{***}$ & $0.116^{***}$\\ 
  & (0.037) & (0.037) & (0.038) & (0.036) \\ 
  & & & \\ 
 $\bar{X}_{i}$ &  0.002$^{***}$& NA & NA & NA \\ 
  & (0.001) &  &  & \\ 
  & & & \\ 
  $X_i(t)$ & NA &  $\hat{\beta}(t)^{***}$ & NA & NA \\ 
  &  &  &  &\\ 
  & & & \\ 
 $Q_i(p)$ & NA & NA & $\hat{\beta}(p)^{***}$ &NA \\ 
  &  &  &  & \\ 
  & & & \\ 
 $Q_{i}(t,p)$ & NA & NA & NA &$\hat{\beta}(t,p)^{***}$ \\ 
  &  &  &  \\ 
  & & & \\ 
 
\hline \\[-1.8ex] 
Observations & 92 & 92 & 92 &92\\ 
Adjusted R$^{2}$ & 0.337 & 0.341 & 0.347 & 0.411\\ 
\hline 
\hline \\[-1.8ex] 
\textit{Note:}  & \multicolumn{3}{r}{$^{*}$p$<$0.1; $^{**}$p$<$0.05; $^{***}$p$<$0.01} \\ 
\end{tabular} 
\end{table}

\section{Supplementary Figures}
%plot for VM and EF
\begin{figure}[H]
\begin{center}
\begin{tabular}{ll}
\includegraphics[width=.45\linewidth , height=.45\linewidth]{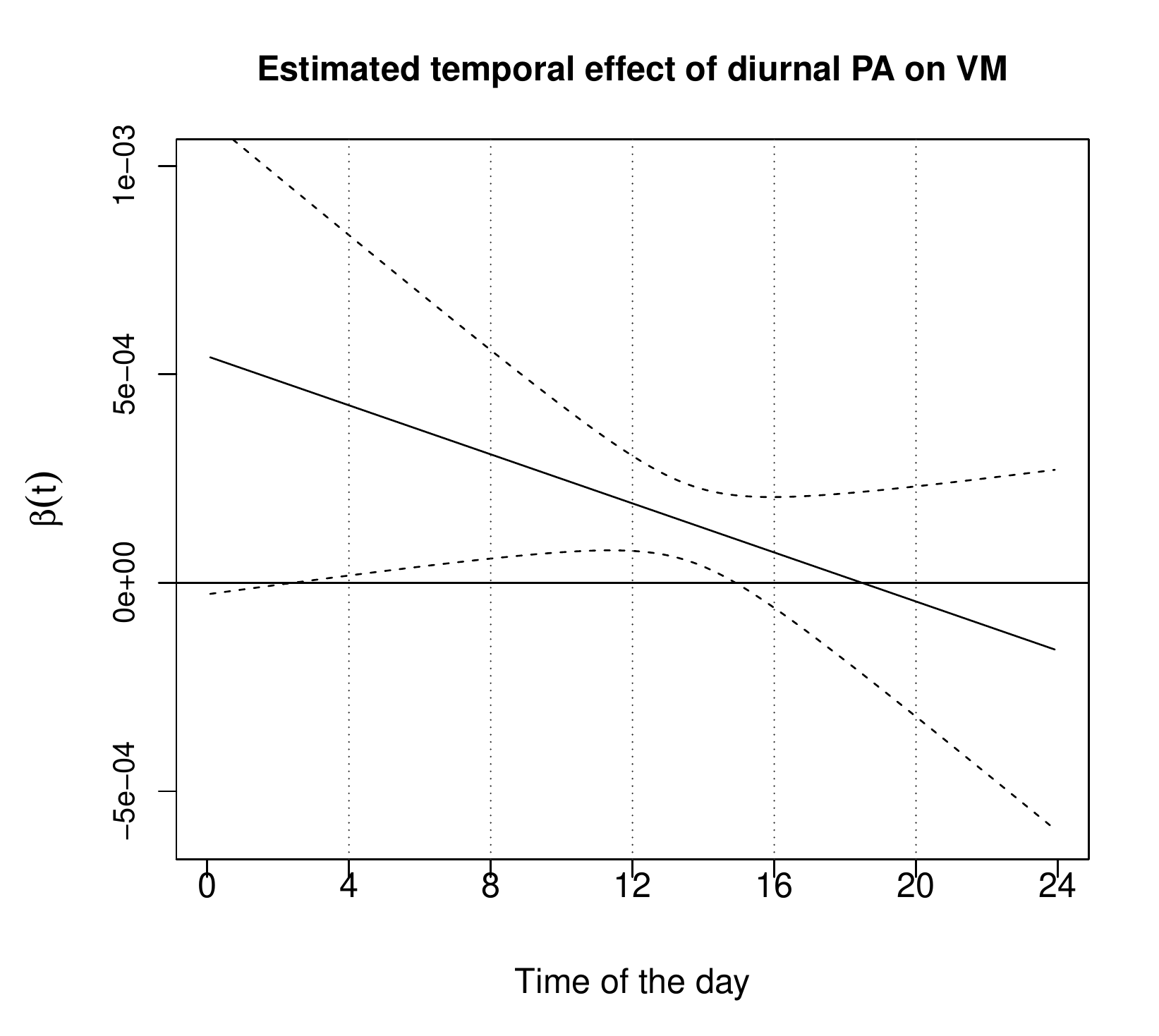} &
\includegraphics[width=.45\linewidth , height=.45\linewidth]{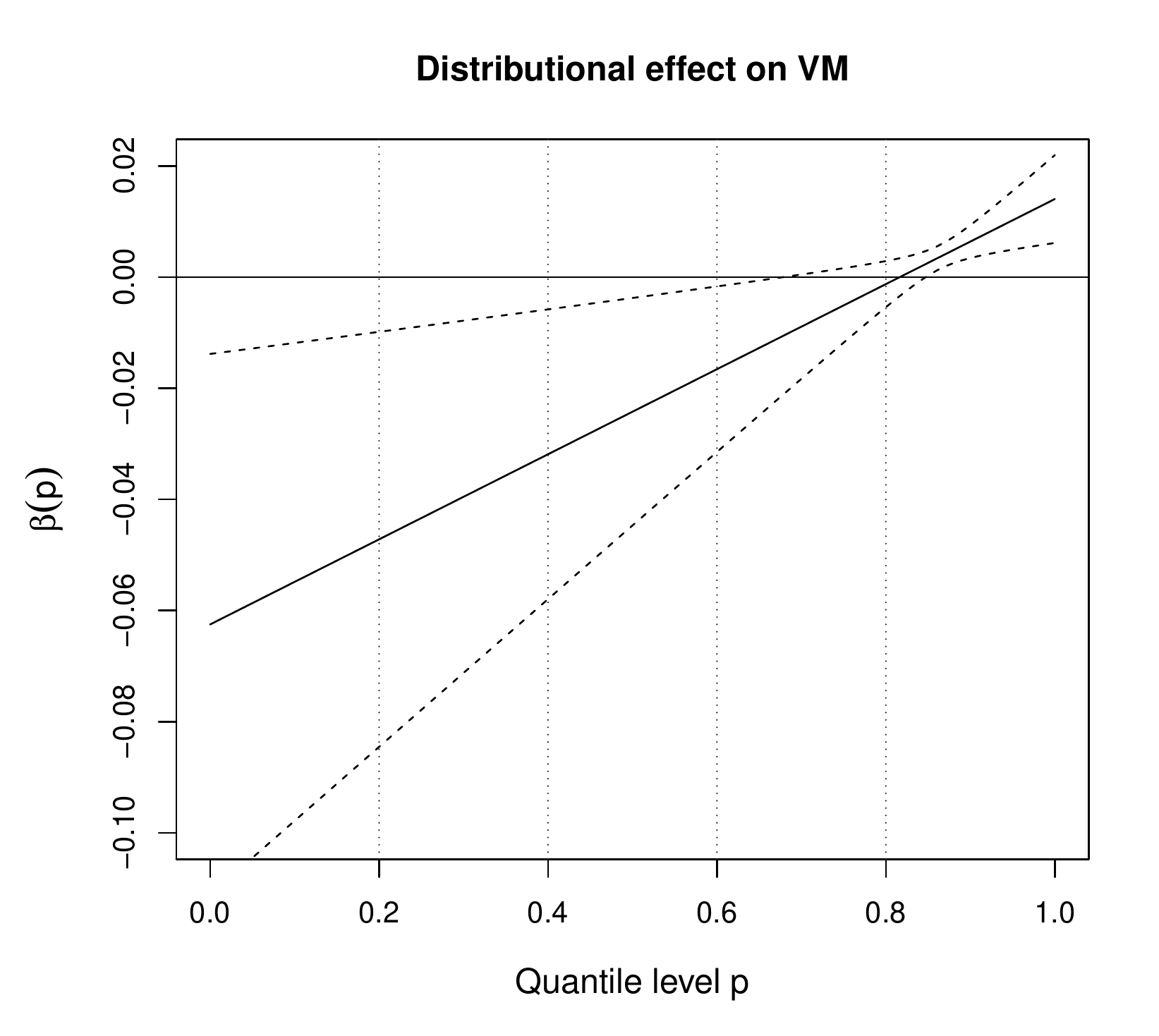}\\
\includegraphics[width=.51\linewidth , height=.55\linewidth]{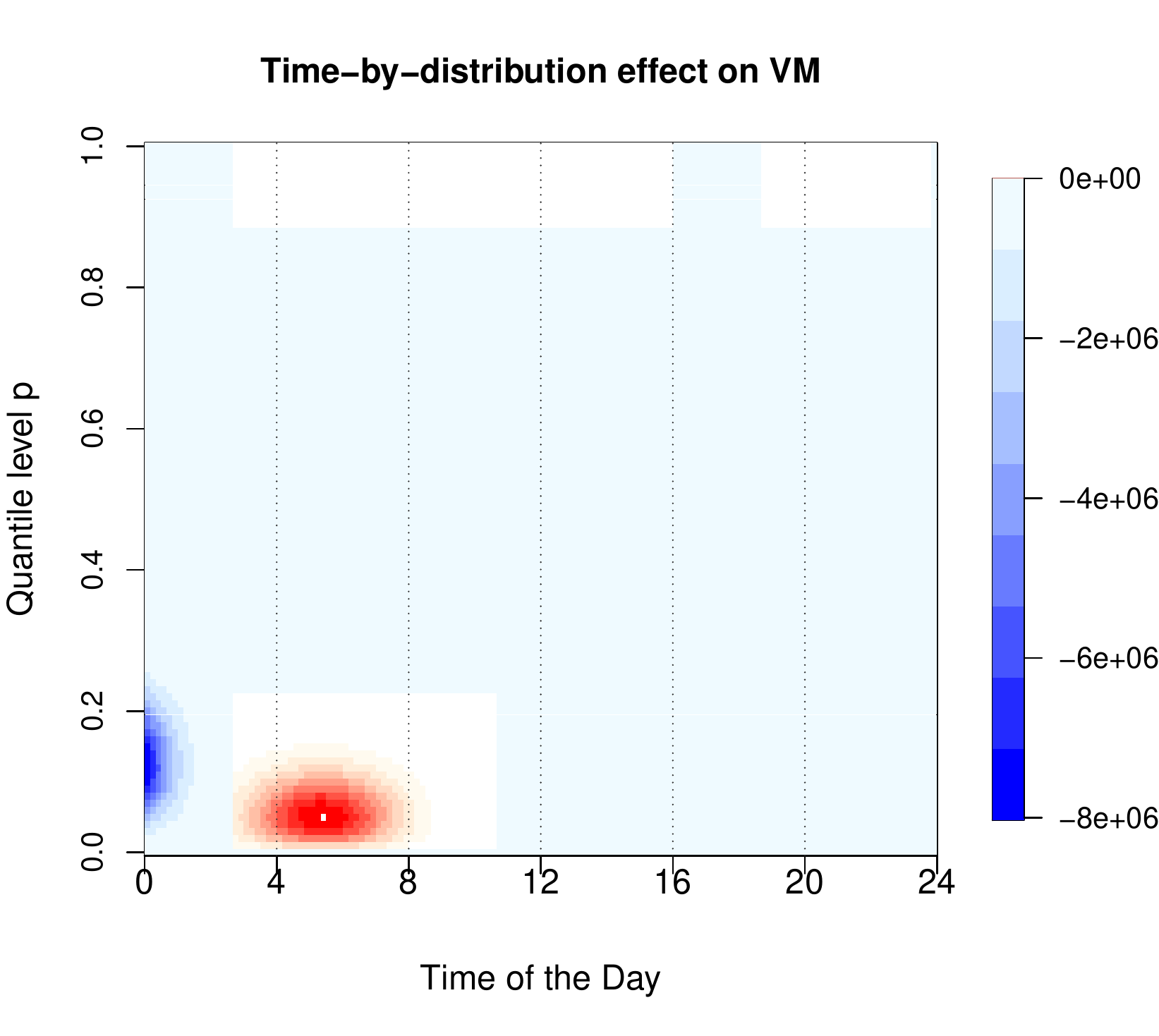}& 
\includegraphics[width=.51\linewidth , height=.55\linewidth]{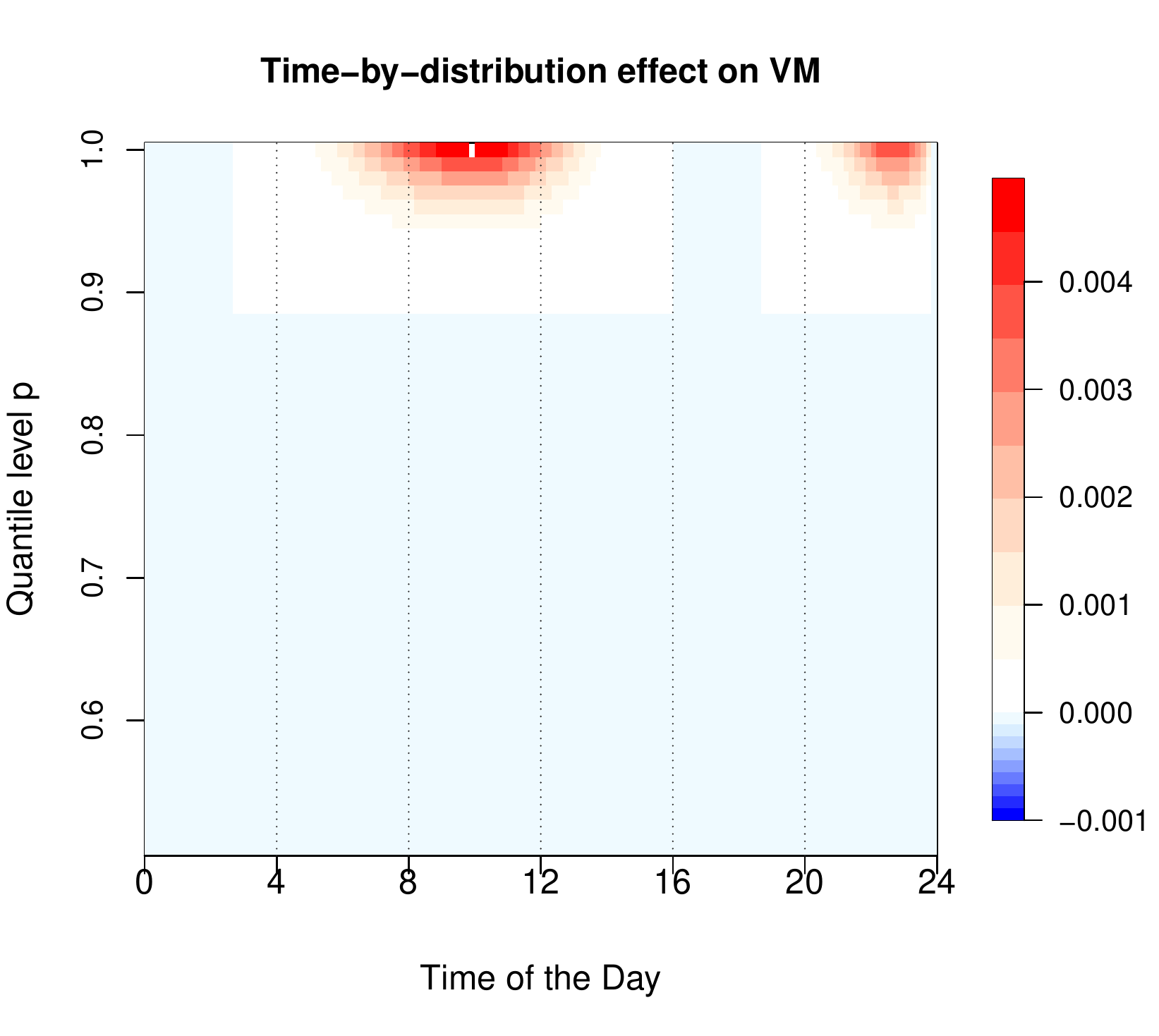}
\\
\end{tabular}
\end{center}
\caption{The estimated effects of the different PA metrics (Model 2-4) on VM score. Estimated temporal effect  (solid line) $\beta(t)$ (top left). Estimated distributional effect $\beta(p)$ (top right). Estimated bivariate effect $\beta(t,p)$ of time-by-distribution PA surface (bottomleft). The same plot with $p$ restricted to $(0.5,1)$ (bottomright). Higher maximal PA during the morning and night are found be associated with a higher score of VM. Higher minimal PA during early morning also appears to be associated with higher score of VM, caution should be taken when interpreting the results for $p<0.5$.}
\label{fig:fig5comb}
\end{figure}

\begin{figure}[H]
\begin{center}
\begin{tabular}{ll}
\includegraphics[width=.45\linewidth , height=.45\linewidth]{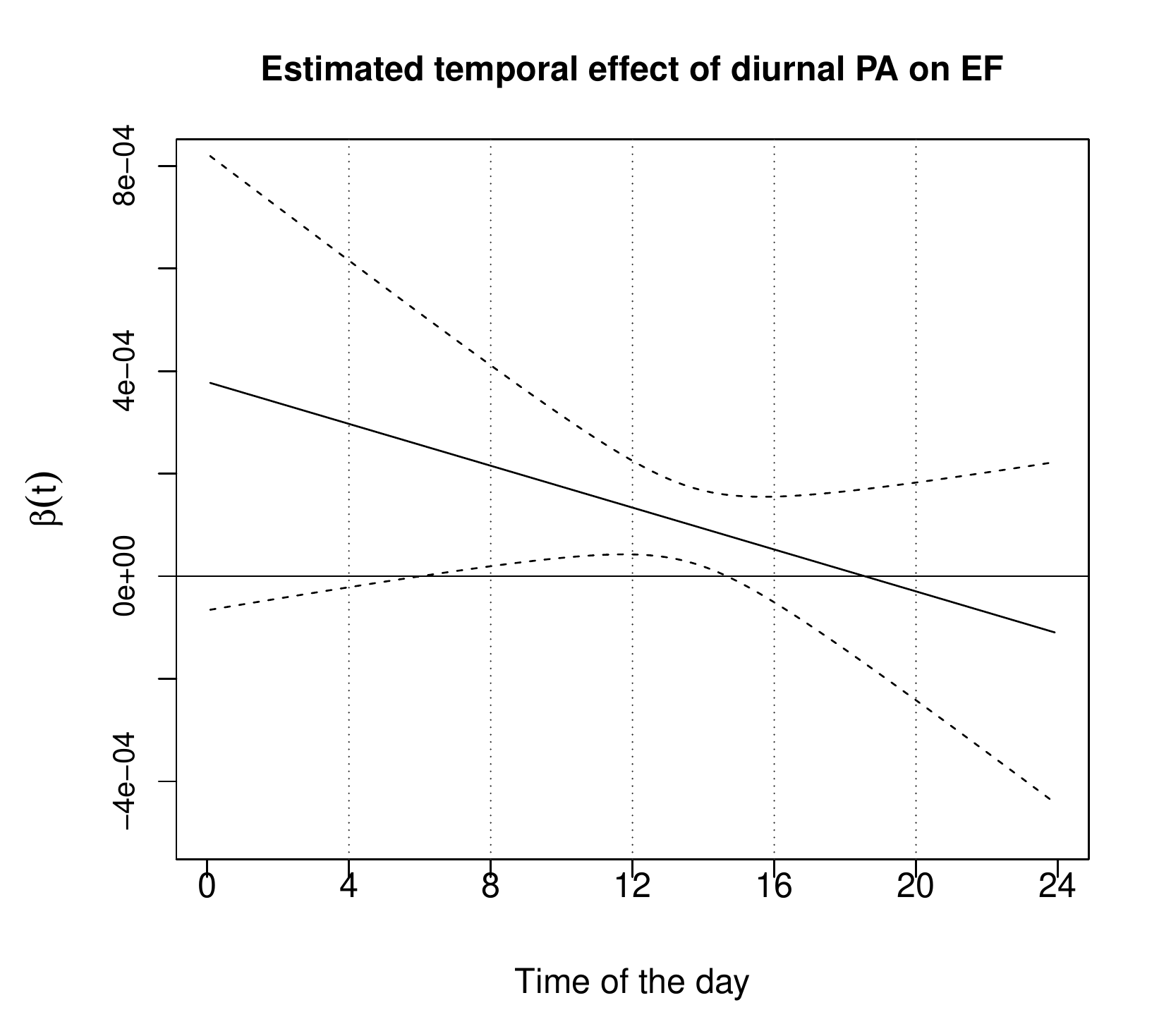} &
\includegraphics[width=.45\linewidth , height=.45\linewidth]{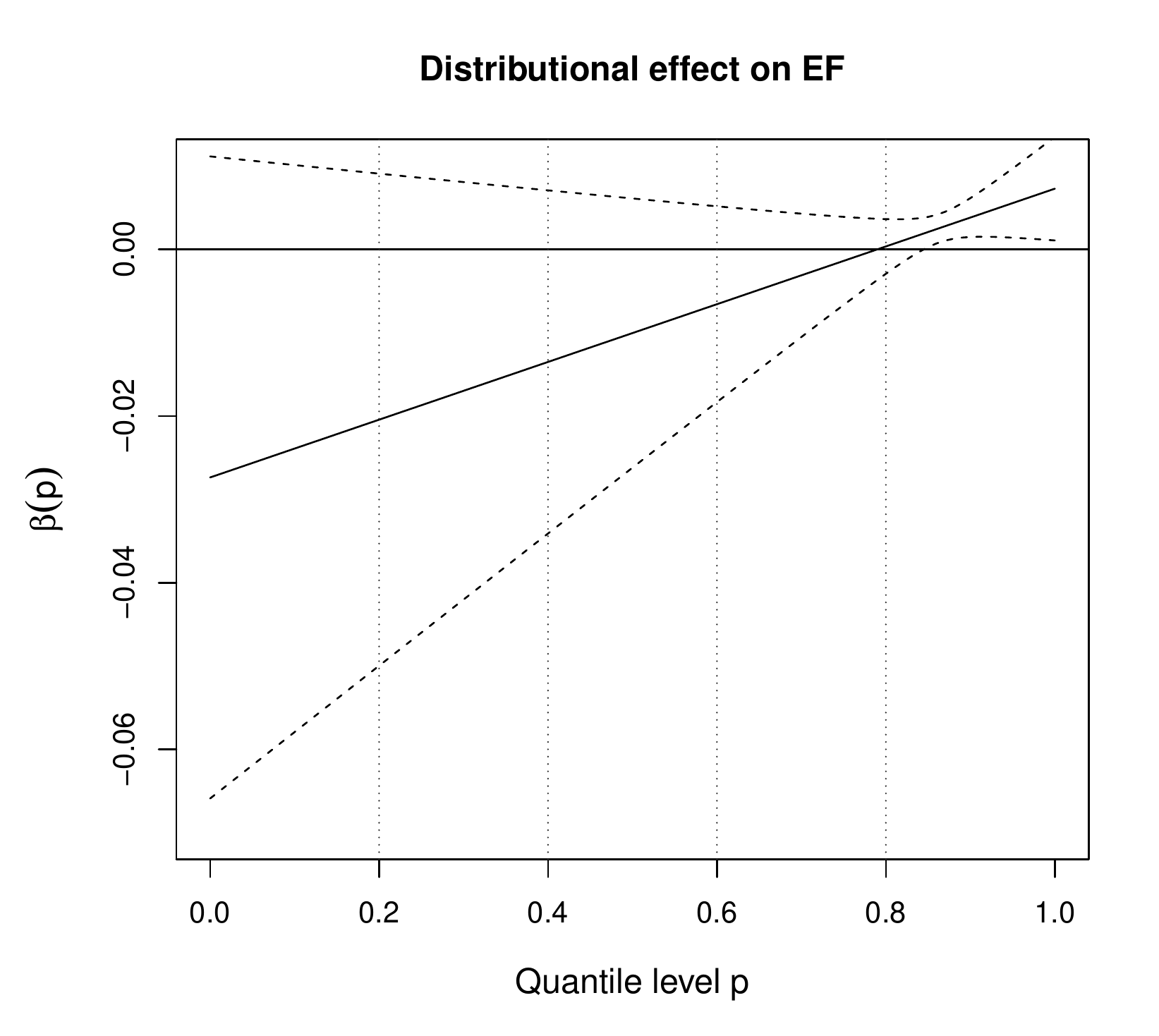}\\
\includegraphics[width=.51\linewidth , height=.55\linewidth]{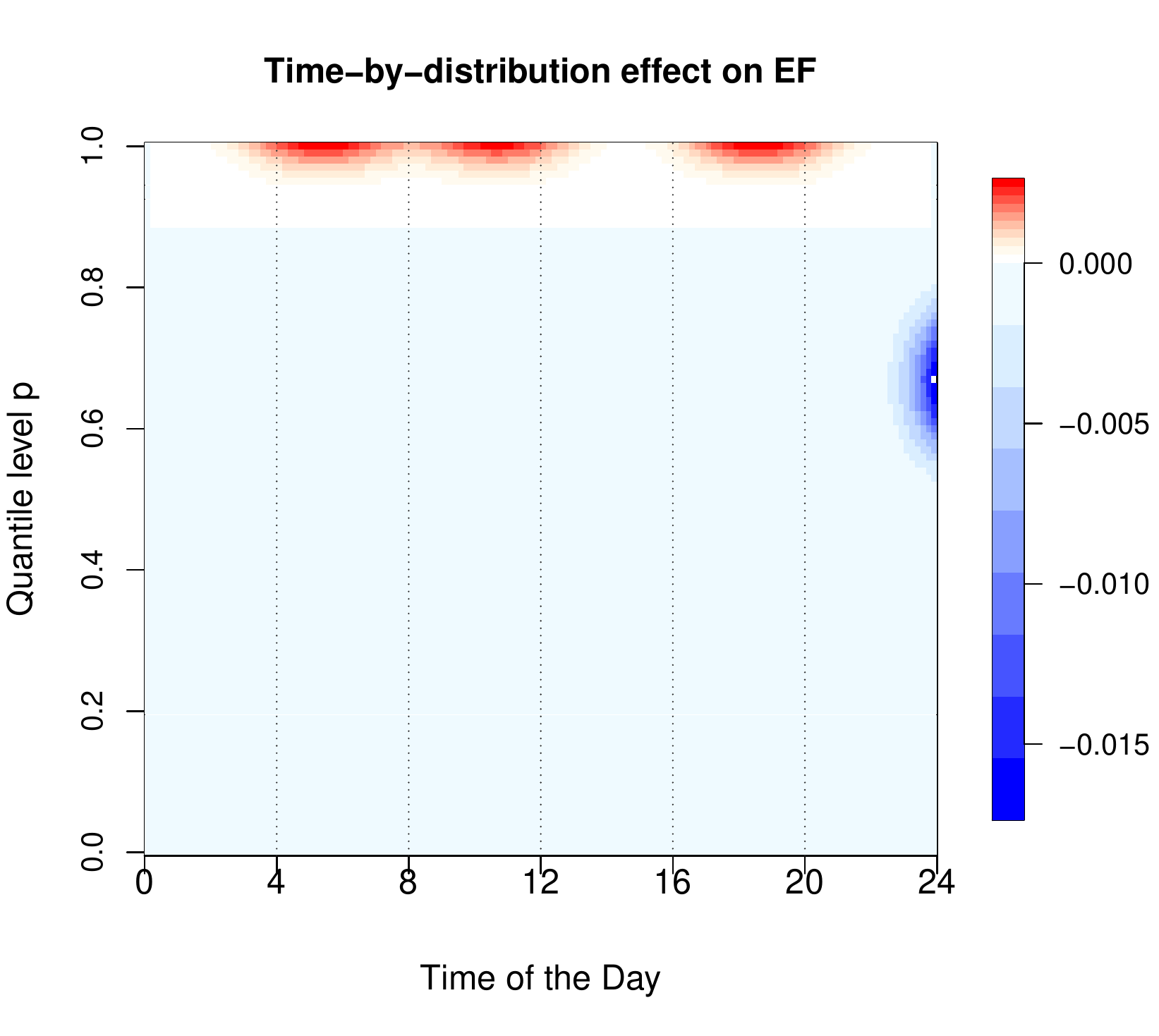}& 

\\
\end{tabular}
\end{center}
\caption{The estimated effects of the different PA metrics (Model 2-4) on EF score. Estimated temporal effect  (solid line) $\beta(t)$ (top left). Estimated distributional effect $\beta(p)$ (top right). Estimated bivariate effect $\beta(t,p)$ of time-by-distribution PA surface (bottomleft). Higher maximal PA during the morning and evening is found to be associated with higher EF score.}
\label{fig:fig6comb}
\end{figure}

\end{document}